\documentclass[twocolumn]{aastex6}
\usepackage{graphicx}
\usepackage{epstopdf}
\usepackage{natbib}
\usepackage{rotating}

\def\be{\begin{equation}}
\def\ee{\end{equation}}
\def\bea{\begin{eqnarray}}
\def\eea{\end{eqnarray}}
 
\newcommand{\HI}{\hbox{{\rm H}\kern 0.1em{\sc i}}}
\newcommand{\HII}{\hbox{{\rm H}\kern 0.1em{\sc ii}}}
\newcommand{\NII}{\hbox{{\rm N}\kern 0.1em{\sc ii}}}

\begin{document}

\title{HALOGAS Observations of NGC 4559: Anomalous and Extra-Planar {\HI} and its Relation to Star Formation}

\author{Carlos J. Vargas\altaffilmark{1}, George Heald\altaffilmark{2,3,4}, Ren\'e A.M. Walterbos\altaffilmark{1}, Filippo Fraternali\altaffilmark{4,5}, Maria T. Patterson\altaffilmark{6}, Richard J. Rand\altaffilmark{7}, Gyula I. G. J\'ozsa\altaffilmark{8,9,10}, Gianfranco Gentile\altaffilmark{11}, Paolo Serra\altaffilmark{12}}

\altaffiltext{1}{Department of Astronomy, New Mexico State University, Las Cruces, NM 88001}
\altaffiltext{2}{CSIRO Astronomy and Space Science, 26 Dick Perry Ave, Kensington WA 6151 Australia}      
      \altaffiltext{3}{The Netherlands Institute for Radio Astronomy (ASTRON), Dwingeloo, The Netherlands}
      \altaffiltext{4}{Kapteyn Astronomical Institute, University of Groningen,
Postbus 800, 9700 AV Groningen, The Netherlands}
      \altaffiltext{5}{Dept. of Physics and Astronomy, University of Bologna, Viale Berti Pichat 6/2, 40127, Bologna, Italy}
      \altaffiltext{6}{Department of Astronomy, University of Washington, Box 351580, Seattle, WA 98195}
      \altaffiltext{7}{Department of Physics and Astronomy, University of New Mexico, 1919 Lomas Blvd. NE, Albuquerque, NM 87131}
	  \altaffiltext{8}{SKA South Africa Radio Astronomy Research Group, 3rd Floor, The Park, Park Road, Pinelands 7405, South Africa}
	  \altaffiltext{9}{Rhodes Centre for Radio Astronomy Techniques \& Technologies, Department of Physics and Electronics, Rhodes University, PO Box 94, Grahamstown 6140, South Africa}     
	  \altaffiltext{10}{Argelander-Institut f{\"u}r Astronomie, Auf dem H{\"u}gel 71, D-53121 Bonn, Germany} 
      \altaffiltext{11}{Department of Physics and Astrophysics, Vrije Universiteit Brussel, Pleinlaan 2, 1050 Brussels, Belgium}
      \altaffiltext{12}{INAF – Osservatorio Astronomico di Cagliari, Via della Scienza 5, I-09047 Selargius (CA), Italy}

\begin{abstract}

We use new deep 21 cm {\HI} observations of the moderately inclined galaxy NGC 4559 in the 
HALOGAS survey to investigate
the properties of extra-planar gas. We use TiRiFiC to construct simulated data cubes to match the {\HI}
observations. We find that a thick disk component of scale height $\sim\,2\,\mathrm{kpc}$, characterized by a negative vertical gradient in its rotation velocity (lag) of $\sim13 \pm 5$ km s$^{-1}$ kpc$^{-1}$ is an adequate fit to extra-planar gas features. The tilted ring models also present evidence for a decrease in the magnitude of the lag outside of $R_{25}$, and a radial inflow of $\sim 10$ km s$^{-1}$. We extracted lagging extra-planar gas through Gaussian velocity profile fitting. From both the 3D models and and extraction analyses we conclude that $\sim10-20\%$ of the total {\HI} mass is extra-planar. 
Most of the extra-planar gas is spatially coincident with regions of star formation in spiral arms, as traced by H$\alpha$ and GALEX FUV images, so it is likely due to star formation processes driving a galactic fountain. We also find the signature of a filament of a kinematically ``forbidden" {\HI}, containing $\sim 1.4\times 10^{6}$ M$_{\odot}$ of {\HI}, and discuss its potential relationship to a nearby {\HI} hole. We discover a previously undetected dwarf galaxy in {\HI} located $\sim 0.4^{\circ}$ ($\sim 58$ kpc) from the center of NGC 4559, containing $\sim 4\times10^{5}$ M$_{\odot}$. This dwarf has counterpart sources in SDSS with spectra typical of {\HII} regions, and we conclude it is two merging blue compact dwarf galaxies.

\end{abstract}

\section{Introduction}

Substantial reservoirs of material have been documented to exist
outside of the plane of disk galaxies \citep{putmanetal12}. This
extra-planar material has been found at multiple wavelengths and emission sources,
including X-rays \citep{tullmannetal06}, dust \citep{howketal99},
H$\alpha$ \citep{rossaetal03}, and {\HI}
\citep{fraternalietal02,oosterlooetal07,healdetal11}, indicating gas
over a wide range in temperatures and densities. Extra-planar material is likely an excellent probe of the effects spiral galaxies have on their environments, and vice versa. 

It is possible that extra-planar matter could have originally been a
part of the disk of the underlying galaxy, but was expelled from the disk through various energetic
processes, such as supernova explosions or stellar winds from massive
stars. This material could then expand, and rain back down onto the galaxy after cooling. This
process is referred to as the galactic fountain mechanism
\citep{shapirofield76}. The rotation velocity of this material about the galactic center is
expected to decrease with distance from the plane \citep{bregman80}. This reduction in
rotation velocity is generally referred to as a ``lag'', and is a
signature of extra-planar matter. However, observations of
extra-planar gas show lag magnitudes that are larger than
could be reproduced with ballistic models, implying that additional
mechanisms are needed to explain the behavior of this material
\citep{collinsetal02,fraternalibinney06}. \citet{marinaccietal11} postulated that
this extra interaction could be with a hot, but slowly rotating corona
of gas already residing above the disks of galaxies. Alternately, a study by \citet{benjamin02} using a hydrostatic disk model finds that pressure gradients or magnetic tension could also play a part in setting the magnitude of observed lags.

Observationally, lagging gas can be found in both edge-on and inclined
galaxies. In edge-on situations, one can directly measure the vertical
extent and lag in the rotational velocities of the extra-planar
gas. In moderately inclined galaxies, the disentangling of
extra-planar gas from disk gas is more difficult, but is possible if a detectable lag exists, as in
\cite{fraternalietal02,barbierietal05,boomsmaetal08}. Additionally, in
such galaxies the connection with star formation across the disks is
easier to establish. 

The Westerbork Hydrogen Accretion in LOcal GAlaxieS (HALOGAS) survey
\citep{healdetal11} targets 22 nearby moderately inclined and edge-on
spiral galaxies for deep 21 cm observations using the Westerbork
Synthesis Radio Telescope (WSRT). This survey has increased the sample
of nearby galaxies for which extra-planar gas can be characterized,
and one of its goals is to search for a connection between extra-planar
{\HI} and star formation, and externally originating {\HI} with
galactic fountain gas. In this paper we present a detailed analysis of
the HALOGAS data cube for the moderately inclined spiral NGC 4559.  The
Hubble type of the galaxy cited in \citet{healdetal11} is SABcd, though a kinematic influence 
on the {\HI} from a bar is not apparent. The
adopted distance to the galaxy, $7.9$ Mpc, was obtained through the
median of all Tully-Fisher distances.

\begin{table}
\centering
    \begin{tabular}{c c}
    \hline
    \hline
    Parameter & Value \\ 
    \hline
    Hubble Type & SABcd \\
    Adopted Distance & 7.9 Mpc ($1\arcsec$ $=$ $38.3$ pc) \\
    $M_B$ & $-20.07$ \\
    $D_{25}$ & $11.3\arcmin$ \\
    $v_{max}$ & $\sim 130$ km s$^{-1}$ \\
    Total Mass (Virial) & $\sim2.8\times10^{11}$ M$_{\odot}$\\
    \hline
\end{tabular}
\caption{Parameters of NGC 4559. The total mass was computed using \citet{klypinetal11}, and represents the virial mass of NGC 4559.}
\end{table}    
   
\begin{table*}
\centering	
	\begin{tabular}{c c c}
	\hline
	\hline	    
	Parameter & Smoothed Cube & Full Resolution Cube\\
	\hline
Synthesized beam FWHM & $30\arcsec$ $\times$ $30\arcsec $ & $28.38\arcsec \times 13.10\arcsec$ \\
    Conversion Factor & $0.673$ K mJy$^{-1}$  & $1.630$ K mJy$^{-1}$\\ 
    Velocity resolution & $4.12$ km s$^{-1}$ & $4.12$ km s$^{-1}$ \\
    RMS noise per channel & $0.24$ mJy beam$^{-1}$; $0.16$ K & $0.20$ mJy beam$^{-1}$; $0.26$ K\\
    Minimum Detectable Column Density ($5\sigma$) & $1.81$ $\times$ $10^{19}$ atoms cm$^{-2}$ & $3.67$ $\times$ $10^{19}$ atoms cm$^{-2}$\\
    \hline   
    \end{tabular}
    \caption{Observational Parameters. The minimum detectable column density is calculated for a source included in 3 channels ($\sim12$ km s$^{-1}$) at 5$\sigma$. The total integration time of the observations spanned $10\times 12$ hours. See \citet{healdetal11} for further discussion of parameters.}    
\end{table*}

A previous {\HI} study by \citet{barbierietal05} (hereafter, B05)
revealed many interesting details about the gas in NGC 4559. The study
found evidence of $\sim5.9 \times 10^8$ M$_{\odot}$ of extra-planar
gas with a scale height of $\sim2$ kpc, rotating $25-50$ km
s$^{-1}$ slower than the uniform thin disk of {\HI}. The
extra-planar gas was found to be kinematically and spatially regular
throughout the galaxy. Though accretion from the intergalactic medium
(IGM) could not be ruled out, the regular extent of extra-planar gas
suggests it is likely due to a widespread phenomenon, such as star
formation across the disk. B05 used 3-D tilted ring models to model the extra-planar gas, where the thick disk had a separate rotation curve from the thin disk.
We build upon this result by constraining the magnitude of vertical lag, rather than computing a separate rotation curve for the extra-planar gas. B05 also found evidence for a large {\HI}
hole at $\alpha$ $=$ $12^h 36^m 4^s$ $\delta$ $=$ $27^\circ 57{\arcmin} 7{\arcsec}$, that would require $\sim 2\times 10^7$
M$_\odot$ of {\HI} to fill. They determined the {\HI} distribution to
be highly asymmetric between the approaching and receding halves of
the galaxy. Interestingly, B05 also found evidence for a
stream of gas located at ``forbidden" velocity in the major axis position-velocity diagram, near the center of the galaxy, which they
postulate to be associated with the aforementioned {\HI} hole. We further the discussion on the possible origins of this forbidden velocity feature and its possible relation to the hole.

The aim of this work is to expand upon the analysis done by B05 using
the more sensitive HALOGAS
observations of the same galaxy, together with ancillary H$\alpha$ and GALEX FUV observations as tracers of star formation activity. We present the data in Sections 2 and
3, explore three-dimensional tilted ring models to characterize the
presence of extra-planar gas in Section 4, and determine the mass and relation to star formation of
the extra-planar gas in Section 5. In the remainder of Section 5 we
further characterize the forbidden gas feature discovered in B05
and discuss its potential origins and connection to the nearby {\HI}
hole. Section 6 discusses a new detection of
{\HI} in a nearby dwarf companion galaxy, while Section 7 concludes the
paper with a discussion of the results.

\section{Data Acquisition and Reduction}

A brief overview of the data collection and reduction process is
included. We refer the reader to \citet{healdetal11} for a
comprehensive description.

We used the HALOGAS survey observations of NGC 4559, taken in the
Maxi-short WSRT configuration, with baselines ranging from 36 m to 2.7
km to maximize sensitivity to faint, extended emission. Nine of the
ten fixed antennas were used with a regular spacing of 144m. The total
bandwidth of 10 MHz was split into 1024 channels (2.06 km s$^{-1}$ per
channel), with two linear polarizations. Galaxy observations spanned
10 $\times$ 12 hr between January and May 2011.

We used Miriad \citep{saultetal95} to perform the data reduction. Data properties are included in Table 2. We
produced multiple data cubes using a variety of weighting schemes. The
cubes were Hanning smoothed to obtain the final velocity resolution of
$4.12$ km s$^{-1}$ per channel, and a beam size of $28.38\arcsec$
$\times$ $13.10\arcsec$. The $1\sigma$ rms noise in a single channel of the full-resolution cube is
 $0.17$ mJy beam$^{-1}$. The minimum detectable column density ($5\sigma$ and $12$ km s$^{-1}$
velocity width) of the full resolution cube is $3.67 \times 10^{19}$ cm$^{-2}$. The widest
field data cube is $1024$ $\times$ $1024$ pixels of size $4\arcsec$, giving it a field of view of
$~68\arcmin$. The galaxy emission is much smaller than that, so we trim the field of view to
$\sim 24\arcmin$ due to data size considerations.

We note the existence of strong solar interference in the data cube of NGC 4559. 
The observations of this galaxy were taken in 10 12-hour blocks with time from January through May, 2011. This timeframe was a period of moderately large solar activity. Five out of the ten tracks were taken in May, with $3.75 - 5.1$ hours of exposed sunlight time. Though the angular separation between the galaxy and the Sun was kept as large as possible ($\sim 120^{\circ}$), there is still solar interference affecting the short baselines. This solar interference was flagged in problematic baselines and timeframes during the data reduction. The flagging reduced inner uv coverage, which makes sensitivity to extended, faint emission lessened. Furthermore, remaining solar interference artifacts preclude the cleaning of the data cube to reach the deepest possible noise level of the HALOGAS data. We attribute this solar interference as the most likely explanation for the lack of appreciable improvement in sensitivity to extended emission over B05. However, we also note that the rms noise per channel is improved by a factor of $\sim 2$ over B05, so we have improved point source sensitivity, despite the lack of extended emission sensitivity.    

To improve sensitivity to faint, extended emission, we smoothed the
original HALOGAS data cube to a $30\arcsec$ $\times$ $30\arcsec$ beam,
making the noise level in a single channel $0.24$ mJy/beam, or $0.16$
K. See Table 2 for details of the $30\arcsec$ $\times$ $30\arcsec$ beam cube. The cube was primary beam corrected using the Miriad task ``linmos'' when calculating total {\HI} mass. We use the $30\arcsec$ $\times$ $30\arcsec$ cube for all tilted ring modelling. Moment maps
were created using the ``moments'' task within the Groningen Image
Processing System (GIPSY; \cite{vanderhulstetal92}). Moment maps were created by first smoothing the original cube to $60\arcsec$ $\times$ $60\arcsec$, with which masks at the $5
\sigma$ level were produced and applied to the full resolution cube.

GALEX FUV and ground-based H$\alpha$ images allow us to investigate in
more detail the correlation between the lagging {\HI} layer and star
formation in NGC 4559. 
The GALEX FUV image from \citet{gildepazetal07}, and a continuum
subtracted H$\alpha$ image are included in our analysis. The H$\alpha$
image was taken by one of the authors (MP) on March 21, 2012, with the Kitt
Peak National Observatory (KPNO) 4-m telescope. The H$\alpha$ exposure
time was $30$ minutes. The mosaic instrument is known to produce artifacts in which bright stars may appear in multiple CCDs. We first removed this crosstalk between CCDs and trimmed the image. The image was bias subtracted, flat fielded with dark sky flats, and stacked from dithered
images. The image was then continuum subtracted with an $10$ minute
exposure $R$-band image taken on the same night. The pixel scale of
the H$\alpha$ image is $0.258\arcsec$ per pixel.

\section{{\HI} Mass and Extent of the {\HI} Disk}

We estimate the total {\HI} mass using the primary beam corrected and masked total {\HI} map and assume the emission is optically thin. We use the standard conversion to column density, where the beam HPBW major and minor axes, a and b respectively, are in units of arcseconds: N$(\rm{cm}^{-2})=1.104 \times 10^{21} \cdot \rm{F}(\rm{mJy}/\rm{beam} \cdot \rm{km}/\rm{s})/(a\times b)$.
The total {\HI} mass obtained from the HALOGAS data cube is in very good agreement with that found in B05
when the same distance is assumed: $4.53 \times 10^9$ M$_{\odot}$ versus $4.48 \times 10^9$ M$_{\odot}$ in B05 using our assumed distance of $7.9$ Mpc. As mentioned in B05, this agrees with alternate and single dish measurements of the mass from \citet{broeils92} and \citet{shostak75}.
This shows that the increase in integration time and sensitivity of the HALOGAS observations
did not discover a larger amount of {\HI} that the observations of B05 were not sensitive
enough to probe. 

\begin{figure*}
\centering
\includegraphics[scale=0.45]{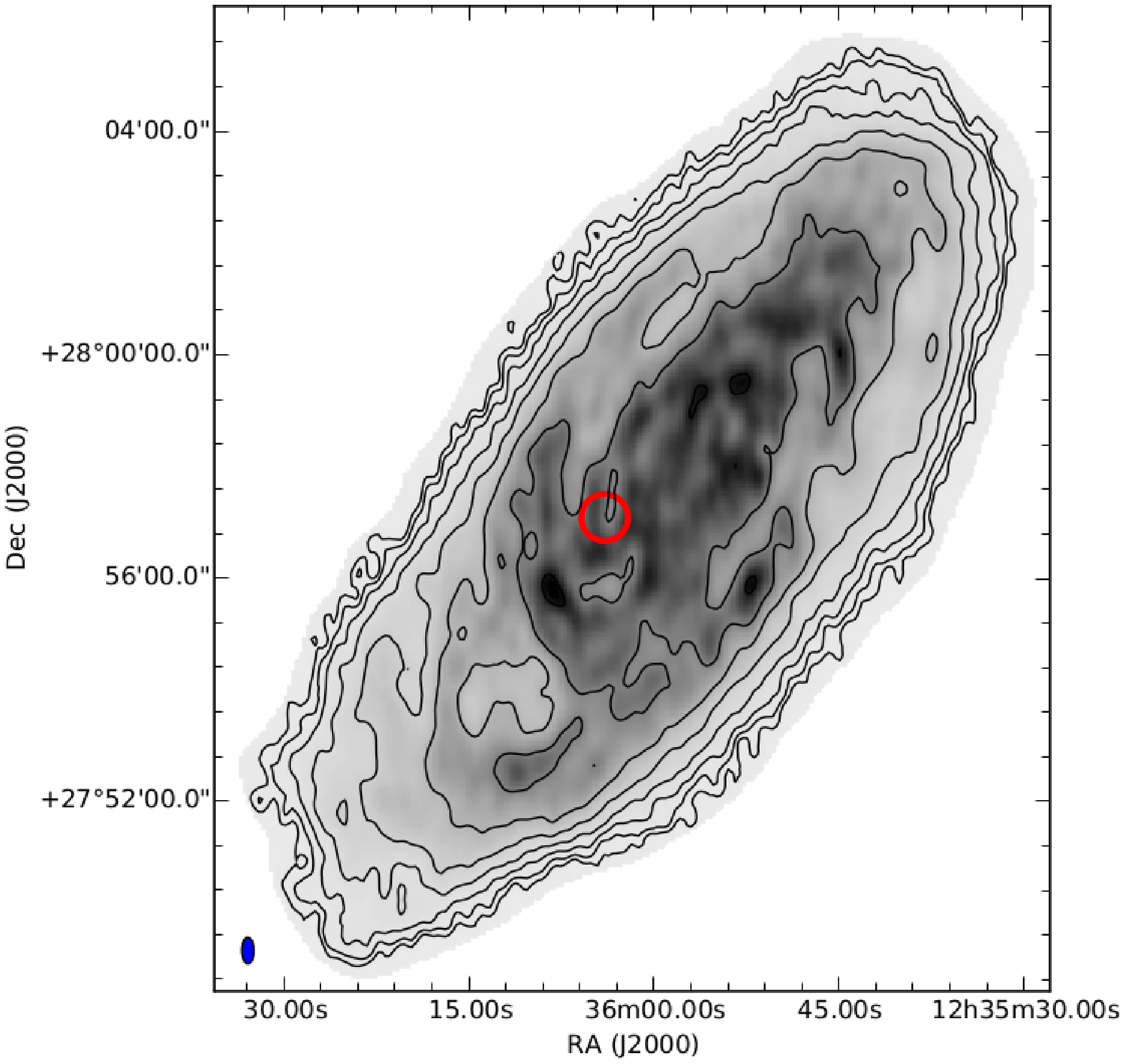}
\includegraphics[scale=0.5]{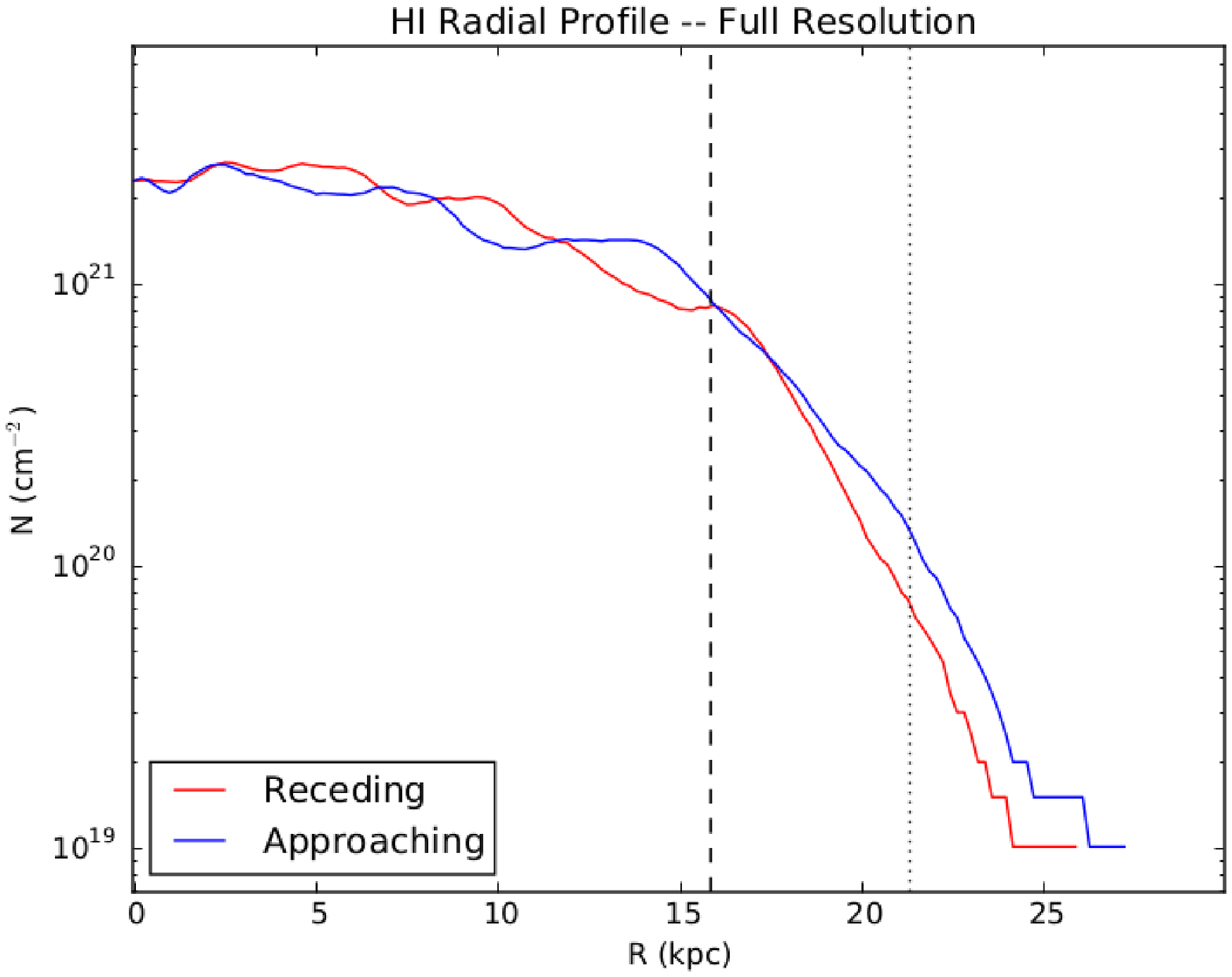}
\caption{Left: Integrated {\HI} map of the highest resolution HALOGAS data cube for NGC 4559. Column density contours begin at $2.4$ $\times$ $10^{19}$ atoms cm$^{-2}$ and increase in multiples of $2$. The red circle marks the location of an {\HI} hole discussed further in Section 5.2. Right: Radial profile obtained from the highest resolution integrated {\HI} map. The vertical dashed line represents R$_{25}$ of the galaxy, and the dotted vertical line is the largest radius at which B05 detects {\HI}, as quoted in that text, and adjusted to our assumed distance of 7.9 Mpc. }
\label{FRmom0}
\end{figure*}

We analyze the extent of the {\HI} disk and radial profile in the highest resolution data cube with a beam size of $28.38\arcsec$ $\times$ $13.10\arcsec$. 
The masked, high resolution cube is integrated over the velocity axis to produce the total {\HI} image, or integrated {\HI} map shown in the top panel of Figure \ref{FRmom0}. We use the GIPSY task \textit{ELLINT} to create an azimuthally averaged surface brightness profiles for the receding and approaching halves independently. \textit{ELLINT} is a 2D ring fitting code that uses a least-squares fitting algorithm to constrain the {\HI} density profile. We provide the moment 0 map as input to \textit{ELLINT} and fit only the surface brightness profile in the approaching and receding sides, independently. We fix the position angle, inclination, and central position using the quoted values in Table 1 of B05 ($-37.0^\circ$, $67.2^\circ$, $\alpha (\mathrm{J2000}) =12^h 35^m 58^s$, and $\delta(\mathrm{J2000}) = 27^\circ 57\arcmin 32\arcsec$ respectively) for the \textit{ELLINT} calculation of the surface brightness profile. The inclination and position angle were both derived from the morphology and kinematics of the tilted ring model analysis done in B05. The central position was also tabulated by B05 using the kinematics of the {\HI} data from B05, and represents the kinematical center of NGC 4559. This is then converted into the column density profiles shown also in Figure \ref{FRmom0}.

The HALOGAS data is somewhat deeper than the data from B05. In Figure \ref{FRmom0}, the radial extent is similar to that found in B05. We calculate the minimum detectable column density in the $30\arcsec$ smoothed cube to be $1.81 \times 10^{19}$ cm$^{-2}$ (see Table 2), which is $1.6$ times lower than the minimum detected column density from the $26\arcsec$ cube in B05: $3.0\times 10^{19}$ cm$^{-2}$. Also, had B05 smoothed their $26\arcsec$ cube to $30\arcsec$, this difference would be slightly smaller. Furthermore, when the same distance is assumed, the HALOGAS data produced an extremely similar total {\HI} mass to B05, implying there is not much extra diffuse {\HI} in the HALOGAS data that was not captured by B05. If we assume there exists a plateau of {\HI} at our limiting column density due to the solar RFI ($1 \times 10^{19}$ atoms cm$^{-2}$) ranging a radial distance between $25-30$ kpc from the center of the galaxy, there would be only $\sim 7 \times 10^{7}$ M$_{\odot}$ of extra {\HI} to detect. 

We do not detect a sharp cut-off in {\HI} but see a rather
constant slope in log (column density) down to the last detected point at $\sim1 \times10^{19}$ atoms cm$^{-2}$.
A cutoff might be expected due the intergalactic ionizing radiation
field, as was found in M83 \citep{healdetal16}. However, this effect would appear near a column density of a
few times $10^{19}$ atoms cm$^{-2}$ (e.g. \citet{corbelli93},
\citet{maloney93}), which is near our sensitivity limit. We also
note the clear change in {\HI} profile morphology inside and outside of R$_{25}$. 
Within R$_{25}$ the {\HI} profile seems clumpy and oscillates, perhaps due to overdensities like spiral arms. Outside of R$_{25}$ the {\HI} distribution becomes more uniform. 

Given the high theoretical sensitivity of the HALOGAS
data cube, it is surprising that the {\HI} column density radial
profile does not reach fainter levels. We suspect this is due to the effects of 
solar interference on the observations (see Section 2). The deconvolution process
of the HALOGAS data is imperfect and cannot fully recover the lack of
short-spacings. In particular, we can detect slightly negative and positive residuals due to solar interference on large angular scales in individual cleaned channel maps at the
level of $\sim 1\times 10^{18}$ atoms cm$^{-2}$. These
residuals change in depth and location between channel maps and
the summation of channels will then limit the sensitivity in the
integrated {\HI} map. If these residuals are summed over $\sim 3$ channels at $3 \sigma$, this limiting column density would approach $\sim 1\times 10^{19}$ atoms cm$^{-2}$, the lowest column densities we observe in the integrated {\HI} radial profile. 
It is likely that a combination of the HALOGAS data with Green Bank Telescope (GBT) observations of NGC 4559 would lead to detection of lower column densities. An effort along these lines is underway using observations with the GBT (Pingel et al. in prep). 

\section{Tilted Ring Models}

\begin{figure*}
\centering
\includegraphics[height=7.6in]{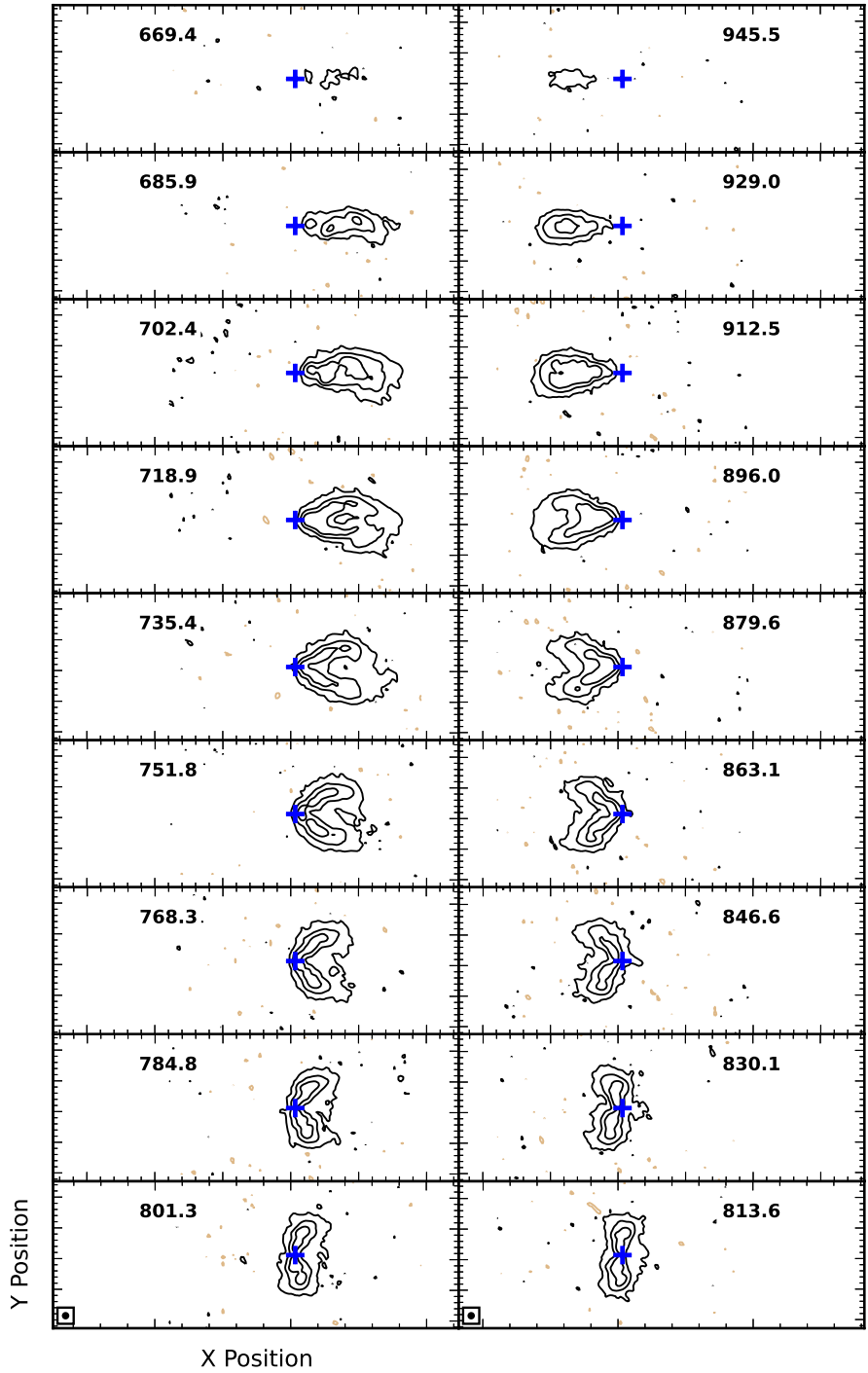}
\caption{ Channel maps for the HALOGAS data cube smoothed to a
  $30\arcsec$ $\times$ $30\arcsec$ beam and rotated $123^\circ$ to make the major-axis horizontal. The kinematic center of the galaxy is marked with a blue cross and the velocity of each channel, in $km/s$, is included. In this plot and all following channel map plots, the right and left hand columns are the receding and approaching halves, respectively. Contour levels are $-3.03$, $3.03$, $27.3$, and $106.1$ ($\times10^{18}$ cm$^{-2}$). The negative contour is colored light brown. Large tick marks are spaced by $1\arcmin$ and $3\arcmin$ in the X and Y axes, respectively.}
\label{chanmaps_data}
\end{figure*}

There are various signs of vertically extended lagging gas within the HALOGAS data cube, itself.
In Figure \ref{chanmaps_data}, we show the channel maps of the $30\arcsec$ resolution HALOGAS
cube, rotated so the major axis is horizontal. Signs of lagging extra-planar gas can be seen as emission that fills the 	``C"-shaped channel maps at intermediate velocities. 

\begin{figure*}
\includegraphics[scale=0.6]{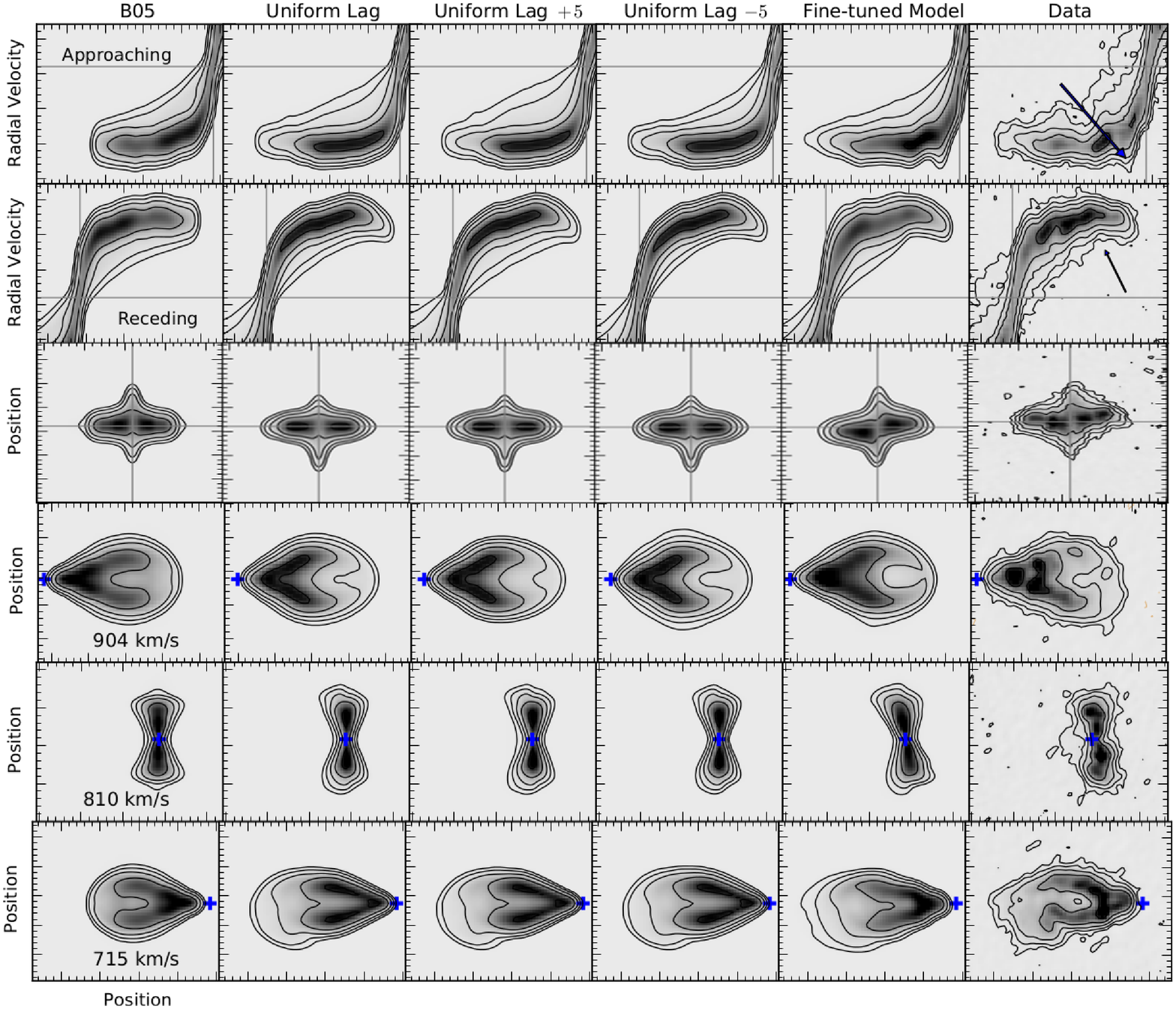}
\caption{ All 3D tilted ring models and data. First row: position-velocity diagram along the major axis zoomed in on the lower velocity approaching side. Second row: position-velocity diagram along the major axis zoomed in on the higher velocity receding side. Third row: position-velocity diagram along the minor axis. Fourth through sixth rows: channel maps at printed velocity in the first column panels. All channel maps are rotated to have the major axis horizontal. All panels have the same contour levels, which begin at $2\sigma$ and increase in multiples of $3$, where $\sigma = 0.24$ mJy beam$^{-1}$. Negative contours are plotted for the data in light brown at $-2\sigma$. Large x-axis tick marks in the top three rows are separated by $8$ kpc, and the y-axis tick marks are separated by $50$ km s$^{-1}$. In the channel maps, large ticks are separated by $7$ kpc. The arrows in the position-velocity diagrams along the major axis mark regions of specific interest when fitting radially dependant velocity dispersion and lag magnitudes (see Section 4). The vertical and horizontal gray lines in the top three rows mark the central position and systemic velocity. The blue $+$ marks in the channel maps mark the kinematic center of the galaxy. }
\label{model_plot}
\end{figure*}

Since the analysis in B05 was done, the vertical velocity structure was measured in edge-on galaxies, like NGC 891 \citep{kamphuisetal07,oosterlooetal07}. Such studies have found that the lagging component is characterized by a vertical gradient in velocity, rather than a bulk decrease in velocity with separate rotation curve from the disk, making the velocity gradient the preferred characterization. Thus, in this study, we constrain the magnitude of the velocity gradient. This is an improvement over B05, where a separate rotation curve for the thick disk was used. To accomplish this, we use TiRiFiC \citep{jozsaetal07} to create 3D tilted ring models to match to the HALOGAS data cube. TiRiFiC is a stand-alone program that constructs 3-D simulated data cubes of rotating galaxy disks. In addition to standard capabilities in other tilted ring codes, TiRiFiC allows for the addition of simple radial and vertical inflows and outflows in the construction of the simulated cubes. We present diagnostic position-velocity diagrams and channel maps of the 3D models as compared to the data in Figure \ref{model_plot}.

We reproduce the ``lagging" model from B05 as a point of comparison. This model contains a two-component gas layer with a thin disk of $0.2$ kpc and a thick disk of $2$ kpc. Both disks have separate rotation curves with the exact values as presented in Figure 7 of B05. The radial surface brightness profile was reproduced from Figure 3 of B05, and $10\%$ of the total was put into the thick disk, as was found in B05. The B05 model is included in the left-most column in Figure \ref{model_plot}.

We create new 3D models to match to the HALOGAS data cube. We use the GIPSY tasks \textit{ELLINT} and \textit{ROTCUR} to find initial estimates for the surface brightness profile and rotation curve. Both tasks are 2D ring fitting codes that use a least-squares fitting algorithm to constrain the {\HI} density profile (\textit{ELLINT}) and the rotation curve (\textit{ROTCUR}). 

We use \textit{ELLINT} in the same fashion here as for the radial profile calculation in Section 3. In a similar fashion, we provide the moment 1 map (velocity field) as input to \textit{ROTCUR}, and fit only the rotation curve in the approaching and receding halves of the galaxy, independently. In the \textit{ROTCUR} fitting of the rotation curve, we fix the position angle, inclination, central position and systemic velocity using the values quoted in Table 1 of B05.   In both tasks we use $61$ rings, all of thickness $15 \arcsec$. We used the initial output surface brightness profile and rotation curve from \textit{ELLINT} and \textit{ROTCUR} as the initial input parameters to TiRiFiC to produce 3D tilted ring models. In TiRiFiC, the values of inclination angle, central position angle, and central position are the same throughout all models and are $67^{\circ}$, $-37^{\circ}$, and $12^h 35^m 58^s$ $+27^\circ 57\arcmin 32\arcsec$, respectively. The approaching half also seems to show a slight warp at large radii. So, the position angle was lowered, through trial and error, by $4^\circ$ beginning at a radius of $18.3$ kpc, in that half of the galaxy. 

\begin{figure*}
\centering
\includegraphics[scale=0.65]{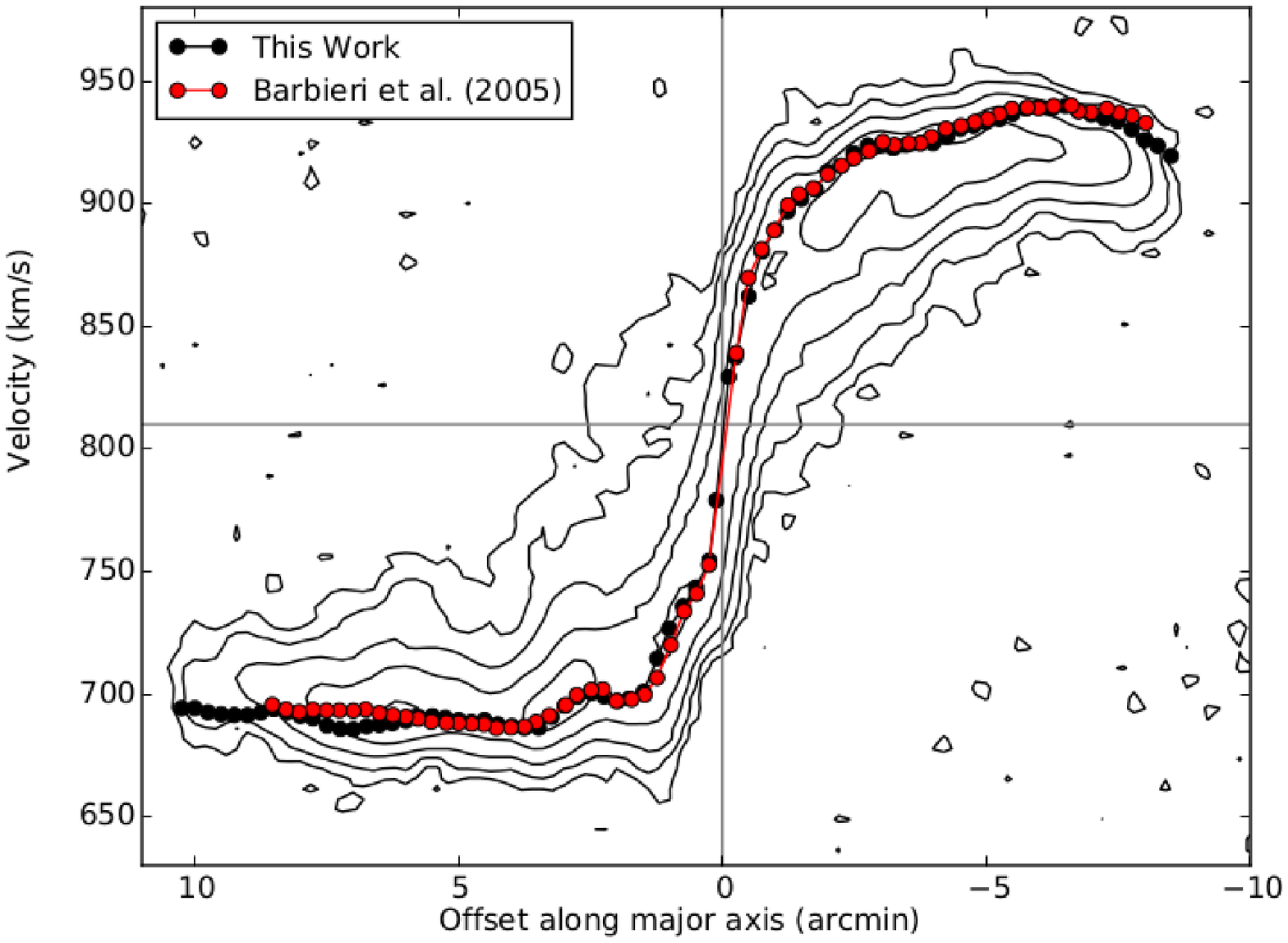}
\caption{Rotation curves of the best fitting thin disk model (black) and the model of 
\citet{barbierietal05} (red) overlaid atop the position-velocity slice along the major axis. Contours begin at $2\sigma$ and increase in multiples of 3. From left to right, this figure moves along the major axis following from south-east to north-west as seen on the sky. Rotation curve values are shown at the inclination of the galaxy. A gas feature at forbidden velocity can be seen in the top-left quadrant and is discussed further in Section 5.2. }
\label{rotcur}
\end{figure*}

\begin{figure}
\centering
\includegraphics[scale=0.48]{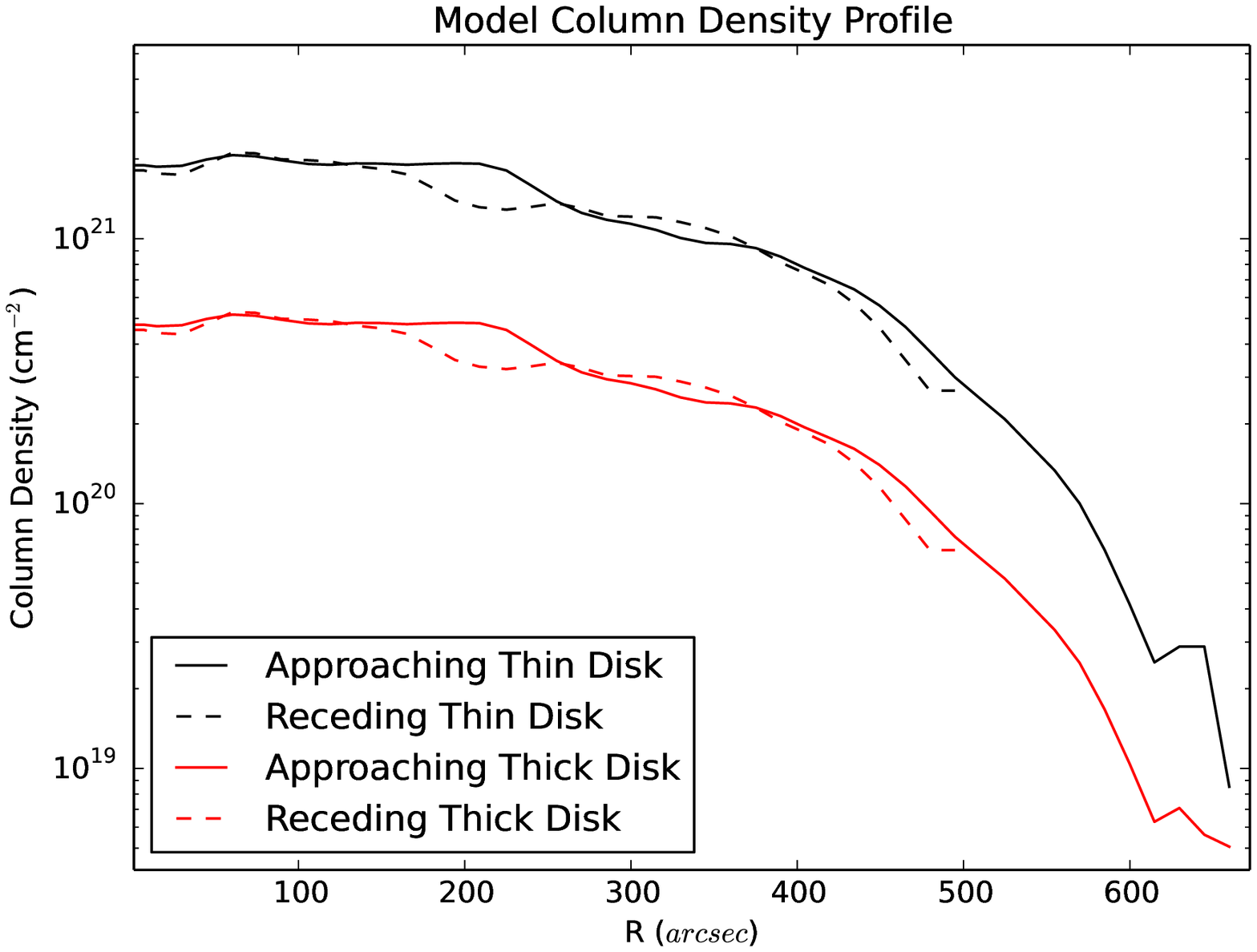}
\caption{Column density profile for each component of the thick disk model. }
\label{coldens_pa}
\end{figure}

Minor adjustments were made to the \textit{ELLINT} and \textit{ROTCUR} output surface brightness profile and rotation curve. These adjustments were made interactively, through trial and error, using TiRiFiC to better match the full 3D structure of the data cube. In all subsequent 3D models, we use trial and error to optimize each parameter by comparing the model to the channel maps, the position-velocity diagrams along both the major and minor axes, the moment 0 map, and the moment 1 map of the $30\arcsec$ $\times$ $30\arcsec$ cube. We do not use TiRiFiC in automated fitting mode, because conventional fitting routines fail to adequately fit faint structures. Since this study is most interested in characterizing faint structures, such as diffuse lagging extra-planar gas and the forbidden gas stream, we elect to fit the cube in this manner. The rotation curve we adopted for all models overlaid on the position-velocity diagram along the major axis and column density profiles are included in Figure \ref{rotcur} and Figure \ref{coldens_pa}.  

We experimented with various warp morphologies by varying the inclination near the edges of the disk, and the scale height in an attempt to reproduce the lagging gas signatures. One such signature can be seen in the position-velocity diagram along the major axis as diffuse emission that is found at velocities closer to systemic than the normally rotating disk. This extra-planar gas feature is commonly referred to as 'beard gas'. No amount of disk warping or disk thickness alone could reproduce the data adequately.
 
We then added a thick disk component to the first models with a uniform lag throughout.  The following parameters were adjusted and matched to the data through trial and error: thick disk scale height, thick disk lag, percentage of total gas in the thick disk, and global velocity dispersion. The velocity dispersion was constrained primarily by matching the thickness and spacing of brightness contours in the position-velocity diagram along the major axis and channel maps. The resulting parameters are shown in Table 2, and the model itself can be seen in the second column of Figure \ref{model_plot}. 

To better illustrate how constrained the key parameter of lag is in this model, we produce two additional models that are identical to the uniform lag model, but with $\pm5$ km s$^{-1}$ kpc$^{-1}$  lag. We note that the lag value is degenerate with other parameters of the model, like the thick disk scale height and velocity dispersion (see Section 4.1).  These two new models are shown in the third and fourth columns of Figure \ref{model_plot}. 

\begin{table*}
\centering

    \begin{tabular}{c c c}
    \hline
    \hline
    Parameter & Uniform Lag Model & Fine-tuned Model \\ 
    \hline
    Thin Disk Scale Height pc & $200$ pc & $200$ pc \\
    Thick Disk Scale Height & $2$ kpc & $2$ kpc \\
    	Central Position Angle & $-37.0^{\circ}$ & $-37.0^{\circ}$ \\
    	Central Inclination & $67.2^{\circ}$ & $67.2^{\circ}$ \\
    	Kinematic Center ($\alpha$, $\delta$, J2000) & $12^h 35^m 58^s$ $+27^{\circ} 57{\arcmin} 32{\arcsec}$  & $12^h 35^m 58^s$ $+27^{\circ} 57{\arcmin} 32{\arcsec}$ \\
    Thick Disk Lag Magnitude & Approaching:$13$; Receding: $6.5$ km s$^{-1}$ kpc$^{-1}$  & Approaching: $13$; Receding: $0-13$ km s$^{-1}$ kpc $^{-1}$\\
    Global Velocity Dispersion & $10$ km s$^{-1}$ & $10$ km s$^{-1}$ + ($0-15$) km  s$^{-1}$ \\
    Percentage Gas in Thick Disk & $20\%$ & $20\%$ \\
    Vertical Density Profile & $sech^2+sech^2$ & $sech^2+sech^2$\\
    \hline   
    \end{tabular}
    
    \caption{Tilted Ring Fitting Parameters of NGC 4559. Central position angle describes the inner, unwarped section of the disk. The fine-tuned model contains all of the same global properties as the Thin + Thick Disk Model, and has variable values for thick disk lag magnitude and velocity dispersion. See the text for details on variable parameters in the fine-tuned model.}    
\end{table*}

In a final model, referred to as the fine-tuned model, we experimented
with various values of velocity dispersion and lag in each individual
ring. 

We converged on a model containing three values of velocity dispersion in three radial extents. For $R \leq 3.4$ kpc, the velocity dispersion is $25$ km s$^{-1}$ in the receding half of the thin disk. For $3.4$ kpc $< R < 12.0$ kpc the dispersion is $18$ km s$^{-1}$. Outside of $12.0$ kpc, the velocity dispersion is $10$ km s$^{-1}$.
This can likely be attributed to turbulent motions in central star forming regions.
In the approaching half, rings at $R = 2.8$ kpc and $R= 3.4$ kpc of the thin disk contain $25$ km s$^{-1}$ of velocity dispersion, to account for the ``bump'' in the position-velocity diagram (see arrow in first row, data column Figure \ref{model_plot}). We
also increase the lag in the receding half to $13$ km s$^{-1}$
kpc$^{-1}$ for $R < 12.0$ kpc of the thick disk, which roughly corresponds with R$_{25}$. The rest of the
thick disk in the receding half of the galaxy contains no lag in this
model due to a sharp cut off in extra-planar gas signatures in the position-velocity diagram along the major axis (see arrow in second row, data column Figure \ref{model_plot}). We note this lack of lag at large radii is not indicative of a sharp radially shallowing lag, but is mostly due to the lack of extra-planar gas at those radii in the receding half. 

The approaching half of the galaxy has a uniform lag throughout in this model, just as in the previous model. Lastly, we include a modest radial inflow along the entirety of the thin disk of $10$ km s$^{-1}$ in order to better match a kink in the position-velocity diagram along the minor axis. We note that inside a radius of 10 kpc, radially inflowing gas with this velocity would reach the center of the galaxy within a Gyr. This model is shown in the fifth column of Figure \ref{model_plot}. 

 The uniform lag model
captures much of the lagging gas component, as seen in the above described extra-planar gas signatures. However, that model requires different lag magnitudes between the approaching and receding halves. Differences between the uniform lag model and that same model with increased and decreased lags (columns two through four of Figure \ref{model_plot}) are most seen in the diffuse signatures of extra-planar lagging gas contours in the position-velocity diagram along the major axis (columns one and two). 

 The fine-tuned model with
its small scale variations in velocity dispersion and lag best represents both halves of the
galaxy. From this analysis it is apparent that the lag magnitude does
not change from one half of the galaxy to the other, but does cut off
at large radii in the receding half, far from the star-forming disk. This result supports a Galactic Fountain model for extra-planar lagging gas, since the lag magnitude is uniform throughout the star-forming disk, and drops off at the edge. Note that no model adequately
represents the forbidden gas. See Section 5.2 for a brief discussion
on modelling the forbidden gas component. Also note that all asymmetries in the fine-tuned model that do not exist in other models arise from the specific treatment of lag, velocity dispersion, and the modest radial inflow that we include exclusively in that model. 

The B05 model is quite comparable to the uniform lag model. We claim that a uniform lag can explain lagging gas just as adequately as two distinct rotation curves for the thin and thick disks in NGC 4559. A uniform lag is preferable in that it has been observationally shown to be more physically accurate \citep{kamphuisetal07}. 

\citet{zschaechneretal15} discuss trends in lags among other galaxies and find that lags seem to reach their radially shallowest values near $R_{25}$. In the fine-tuned model, we model the lag in the receding half of NGC 4559 to cut off sharply at a radius of $12.06$ kpc. At our assumed distance (D$=7.9$ Mpc), $R_{25}$ is $12.98$ kpc, which makes this radial lag cutoff at $0.93R_{25}$. So, this result is consistent with what was found in \citet{zschaechneretal15}. Tilted ring models were created from the HALOGAS data cube of NGC 4559 including radially varying lags, but no radially varying lag producing appreciable improvement to the fine-tuned model. As discussed in \citet{zschaechneretal15}, the overall steepness of lags suggest that conservation of angular momentum is not a sufficient explanation.  Also, \citet{marinaccietal11} create simulations of fountain gas clouds moving through a hot halo medium, and are better at reproducing the steepness of lags seen observationally. Alternatively, \citet{benjamin02} propose that pressure gradients could explain the magnitude of morphology of lags, but this has been difficult to accomplish observationally.
Deep radio continuum observations with the recently upgraded VLA, like \citet{irwinetal12}, should make future measurements of non-thermal pressure gradients possible.  

\subsection{Uncertainties in Derived Parameters}

In general, uncertainties in three-dimensional tilted ring parameters
that describe the data are estimated by varying each individual
parameter to the point where the model no longer adequately represents
the data \citep{gentileetal13}. Certain projections or regions in the
data cube are more sensitive to some parameters than others, so these
intricacies are considered in estimating uncertainties. The decisions
as to what constitutes improvements and acceptable models were based
on visual inspection of the various plots, as in previous papers in
this type of work. The subtle low column density features generally do
not typically lend themselves easily to statistical measures but
visual inspection shows clearly if a certain model feature is required
to reproduce particular faint features in p-v diagrams and channel
maps. 

A comparison between total {\HI} maps is effective at determining the
uncertainty in inclination and position angle. A change of $3^\circ$
in both inclination and position angle through the complete disk is enough to make the total {\HI} maps inconsistent
with the data. 

To constrain the uncertainty in global velocity dispersion, we analyze
position-velocity diagrams. After testing this parameter, the
high velocity contours in the position-velocity diagram along the
major axis no longer represent the data when changed by more than
$\pm3$ km s$^{-1}$ from their original value of $10$ km s$^{-1}$. We note
that the velocity dispersion in the extra-planar gas is degenerate 
with the thick disk scale height and the magnitude of the thick disk lag.
We account for this degeneracy as well as possible in estimating this uncertainty. 

Uncertainties in extra-planar gas related parameters are also
estimated. We assume a thin disk scale height of $200$ pc, while
varying values were used for the thick disk. The fitting done in B05
assumed a thin disk scale height of $200$ pc, so, for consistency, we
retain that value, despite the resolution limits on constraining that
number present in both studies. The central channels in the data cube
are particularly sensitive to scale height changes. We find the thick
disk must have a scale height of $2 \pm 1$ kpc. Since NGC 4559 is not
seen edge on, the relative mass and amplitude of lag for the
extra-planar gas is not easily constrained. We find between
$15\%-25\%$ of the total {\HI} mass in extra-planar gas. We find the
uncertainty in lag to be $\sim 5$ km s$^{-1}$ kpc$^{-1}$ in both halves of the galaxy. These uncertainties take the degeneracy between
scale-height and lag magnitude into account. For instance, large scale
height values can be compensated with small lag magnitudes and vice
versa. However, small imperfections in models, such as location of extra-planar gas signatures in the position-velocity diagram along the major axis, and thickness of diffuse emission in channel maps were closely inspected to minimize this degeneracy.

\section{Anomalous Gas Extraction}

It is useful to separate the lagging extra-planar gas component from
the total {\HI} data cube in an independent way. To that end, we follow
the procedure by \citet{fraternalietal02} to extract the extra-planar
gas component, and create two separate data cubes: one with emission
attributed to regularly rotating gas, and another with only emission
of extra-planar gas. This procedure was also done in B05, so we will focus
on comparing our new results with star-formation tracers.

This procedure assumes that each {\HI} line profile contains a narrow
Gaussian-shaped component whose peak is positioned close to the rotation velocity,
and a broader component whose peak is closer to the systemic velocity. The latter is attributed to extra-planar lagging gas whose
profile's shape is unconstrained, but likely substantially fainter than
the normally rotating component. We estimated the contribution of the normally rotating
component by fitting a Gaussian profile to the upper portions of the
total line profile. Modeling only the tops of the line profiles
enables us to minimize contamination from potential abnormally
rotating components. Experimentation using various percentages of line
profiles was performed to decrease the occurrence of fitting
artifacts. The procedure produced the least artifacts when only the upper $60\%$ of
each line profile was fit. The amplitudes, central velocities, and
widths of the Gaussian profiles were fit
throughout the data cube. Based on experimenting with parameters, a dispersion maximum limit on each Gaussian
profile of $30$ km s$^{-1}$ was imparted on the fitting. The Gaussian
profile fit to each line profile was then subtracted from the data line profiles,
leaving only anomalously rotating extra-planar gas. In $285$ out of $158234$ instances,
Gaussian profiles were not able to be fit. In these
instances, the profiles were excluded from the extra-planar cube.

\begin{figure*}
\centering
\includegraphics[scale=0.8]{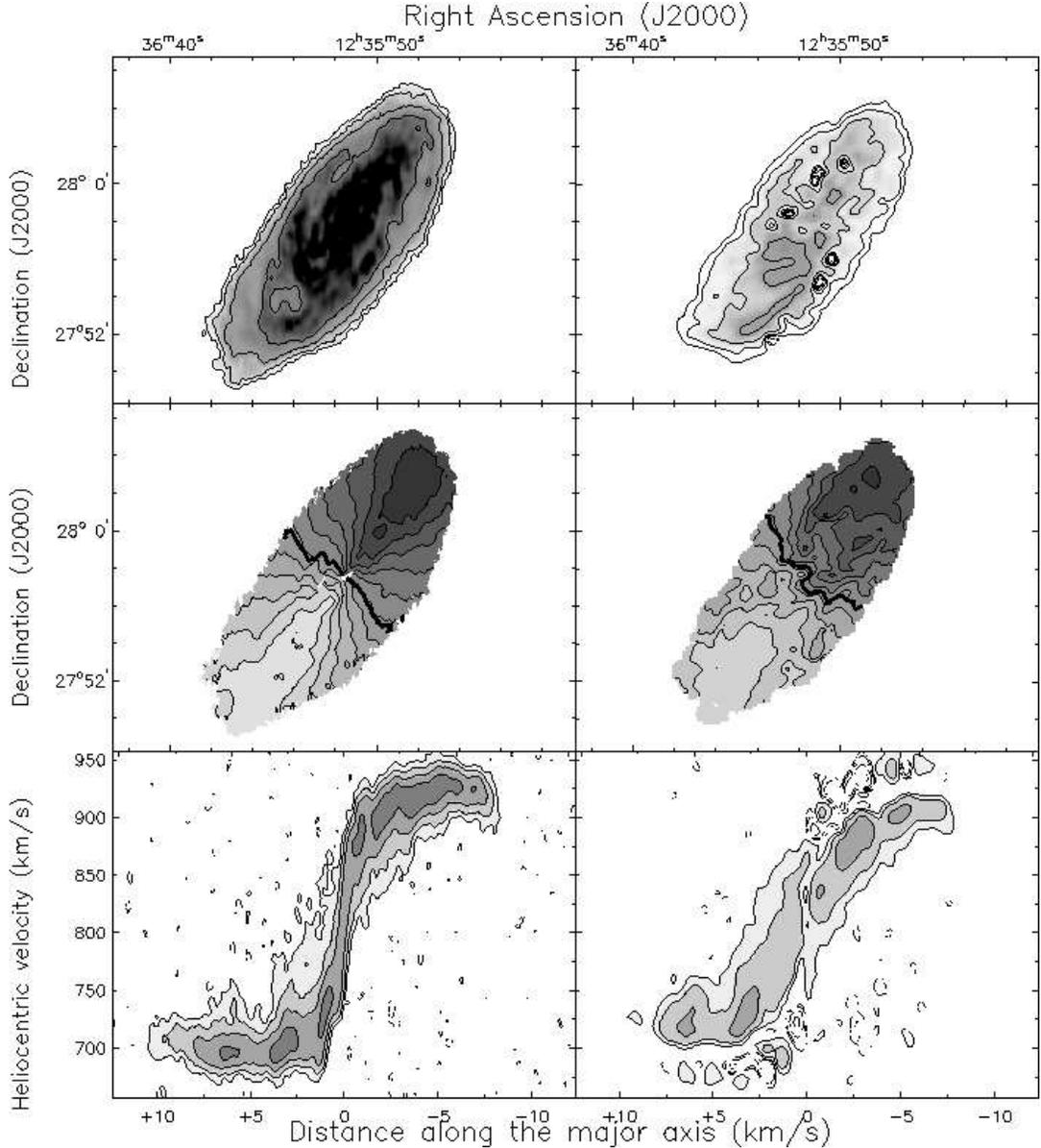}
\caption{Various plots of the extra-planar gas extraction results. Upper left: Integrated {\HI} map of the data, upper right: integrated {\HI} map of the extracted extra-planar gas, middle left: velocity field of data, middle right: velocity field of extracted extra-planar gas, lower left: position-velocity slice along the major axis of the data, lower right: position-velocity slice along the major axis of the extracted extra-planar gas. All column density contours begin at $3\sigma$ in each respective cube and increase in multiples of 3.
The receding half of the galaxy resides to the north-west. The velocity contours in velocity fields range $680$ km s$^{-1}$ to $940$ km s$^{-1}$ in intervals of $20$ km s$^{-1}$. Panels in the left column are at full resolution and panels in the right column are smoothed to $30\arcsec$. The upper x-axis units apply only to the top two rows, while the lower x-axis units apply to the bottom row.}
\label{filippo}
\end{figure*}

\begin{figure*}
\centering
\includegraphics[scale=0.47]{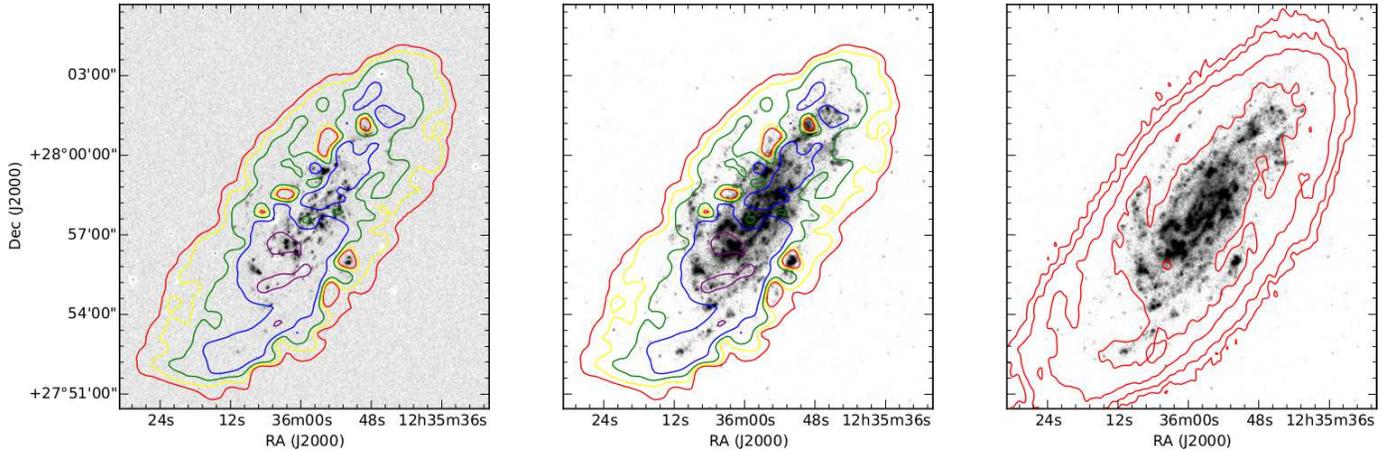}
\caption{Left: KPNO H$\alpha$ image, middle: GALEX FUV image \citep{gildepazetal07}, right: Galex FUV image. Overlaid atop the left two images are the total extracted extra-planar gas moment 0 contours. The colors represent different values of column density and range, from lowest to highest, red, yellow, green, blue, and violet. The contours begin at $3.75\times10^{20}$ atoms cm$^{-2}$ and increase in multiples of $2$. Overlaid atop the rightmost image is the full resolution {\HI} data, the same as Figure \ref{FRmom0}.   }
\label{extrap_mom0}
\end{figure*}

The results of the extra-planar gas extraction can be seen in Figures
\ref{filippo} (right panels) and \ref{extrap_mom0}. The behavior of
the extra-planar gas seen in velocity fields is somewhat more
irregular than in the total data. 
As seen in the bottom right panel of
Figure \ref{filippo}, the residual p-v diagram from the fit (i.e. the presumed
extra-planar gas) includes some gas at extreme values of radial velocity, near $\sim 950$ km s$^{-1}$ and $\sim 675$ km s$^{-1}$, which is
not lagging. This is due to the limitation imposed by the forcing the
fit to only 60\% of the peak of the profile and the maximum cap on the
velocity width of the regularly rotating thin disk in the fit. This gas is also likely due to the assumption of Gaussian velocity profiles. If the gas is clumpy or moving peculiarly the Gaussian profile assumption is incorrect. 
However, most of the residual gas is indeed lagging, as we find $\sim 10\%$ of the total residual to reside in the extreme velocity regimes of the position-velocity diagram along the major axis. 

Assuming the {\HI} emission is optically thin, we can estimate the
total mass in {\HI} from its emission. The total {\HI} mass of the
galaxy is $\sim 4.5 \times 10^{9}$ M$_{\odot}$. The mass of the
extracted extra-planar {\HI} is $\sim 4.0 \times 10^{8}$ M$_{\odot}$,
or $\sim 10\%$ of the total {\HI} mass. The mass in the thick {\HI} disk
model derived from the tilted ring fitting analysis contains 20\% of
the total {\HI} mass. However, the two methods are not directly
comparable since the portion of that thick disk at low z was superimposed over the
thin disk in the ring models. Since the vertical profile used  in the ring modeling was a $sech^2$ model, one
can integrate that function over the thin disk scale height, and exclude that amount of gas in the thick disk that spatially resides within the thin disk. The
amount of thick disk mass outside $\pm3$ times the thin disk scale
height is $14\%$ of the total mass. This is still somewhat more than
for the extracted emission in the Gaussian fitting. In reality, the line profile fitting method certainly misses some extra-planar gas which happens to be at the same velocity (in projection) as the disk gas \citep{fraternaliinprep}. This is particularly true for gas near the minor-axis, where the rotation signal is weak. We conclude that overall the mass estimates for
the extra-planar gas derived from the two methods are in reasonable
agreement, and we find $\sim 10-20\%$ of the total {\HI} mass to be extra-planar.

\subsection{Relation of Extra-planar Gas to Star Formation}

We incorporate two ancillary images -- one H$\alpha$ narrowband image, and one GALEX FUV image from \citet{gildepazetal07} -- as tracers of star formation in NGC 4559.  
We show the extracted extra-planar gas overlaid as colored contours atop the H$\alpha$ image and the GALEX FUV image in Figure \ref{extrap_mom0}.  
In Figure \ref{extrap_mom0}, we see the location of the highest
densities of extra-planar {\HI}. The three highest density contours
(violet, blue and green) trace the regions of active star
formation. Additionally, a spiral arm feature seen extending to the
south-east in both the H$\alpha$ and FUV images is traced by the
extra-planar gas. Note there are some small isolated depressions in
the extracted extra-planar gas. These are not in all cases regions with
lesser amounts of extra-planar gas, but could also be regions
where the Gaussian profile fit did not converge.

\begin{figure*}
\centering
\includegraphics[scale=0.5]{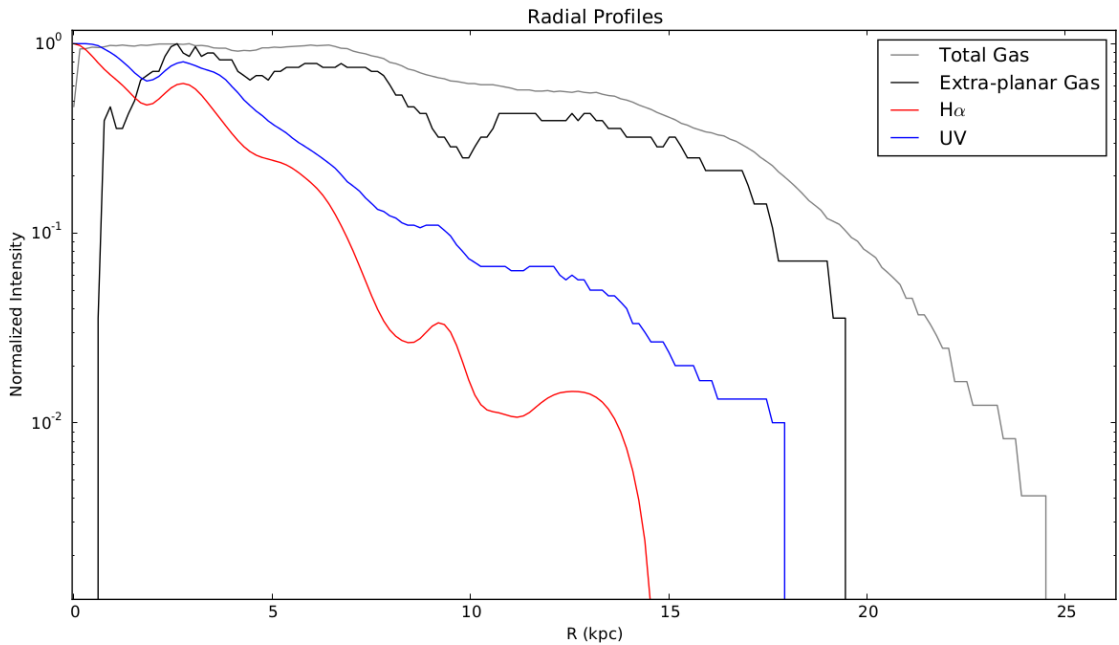}
\caption{Radial profiles obtained from the extracted extra-planar gas, total {\HI} map, H$\alpha$ map, and GALEX UV map. The H$\alpha$ and UV maps are smoothed to match the resolution of the {\HI} map before the radial profiles are calculated. The intensities are normalized to their peak values so differences in profiles can be easily compared. }
\label{radprofs}
\end{figure*}

To better see the radial extent of the extra-planar gas and the star formation tracers, we computed azimuthally averaged radial profiles for the total {\HI}, Gaussian extracted extra-planar gas cube, the H$\alpha$ image, and the GALEX FUV image. All profiles are corrected for the inclination of the galaxy. These radial profiles are shown in Figure \ref{radprofs}. In this figure, the intensity of each component has been normalized to its peak so that differences in structure can be more easily seen. We find that the total {\HI} is more extended than the extra-planar gas. The extra-planar gas traces the extent of the UV profile well. Since UV emission, which is indicative of older star formation, and the extra-planar gas are coincident, it is likely that the extra-planar gas is related to past star formation processes \citep{kennicuttevans12}. Had the extra-planar gas been due to accretion, there would be a high likelihood of seeing more radially extended extra-planar gas that traces the extent of the total {\HI}. Thus, we conclude that the extra-planar gas is most likely due to star formation processes. 

We also note that our modeling showed evidence that the thick disk lag value approached zero outside of R$\sim12$ kpc in the receding half. The modeled thick disk still contained gas there, so this gas would be non-lagging extra-planar gas that would not be found by the Gaussian fitting algorithm. However, we cannot distinguish this gas from disk gas observationally, due to the inclination of the galaxy, so we cannot say whether or not the thick disk lag approaches zero outside of this radius.

It is interesting that the lagging gas is still closely
associated with the spiral arms (see e.g. the southern spiral arm in Figure
\ref{extrap_mom0}). Here, it is relevant to consider the fountain cycle timescale, which is the amount of time it takes for an ejected parcel of gas to fall back down on the disk. If the ejection occurs at $\sim70$ km s$^{-1}$ and is vertically stationary at its apex, then its average vertical velocity is $\sim 35$ km s$^{-1}$. This vertical ejection velocity is consistent with the results of \citet{fraternalibinney06}, where a vertical kick velocity of this magnitude was used in dynamical models matching observed extra-planar properties of NGC 891 and NGC2403. If the gas reaches a vertical height of $\sim 2$ kpc, then the gas will traverse the full vertical extent of $4$ kpc and fall back to the disk in $\sim 100$ Myr. When the gas is at a vertical height of $2$ kpc it experiences a lag of $26$ km s$^{-1}$, and while it is at $1$ kpc it lags the disk by $13$ km s$^{-1}$. Thus, the average magnitude of the lag is $13$ km s$^{-1}$, which would produce an azimuthal offset of $\sim 1.3$ kpc over the course of the $100$ Myr fountain cycle timescale. This is slightly larger than the $30\arcsec$ ($1.1 kpc$) smoothed beam, which is most sensitive to diffuse lagging gas features. Although the inclination of the galaxy would make this offset slightly smaller, we see no evidence for a systematic offset of $0.5-1$ beam. The time scales suggest that the extra-planar gas is likely to be recently ejected fountain gas, that has not had time to cool and begin its journey back down to the disk. So, we conclude that the most likely origin for the bulk of extra-planar {\HI} in NGC 4559 is the galactic fountain mechanism; i.e., {\HI} is transported above the disk as cold/warm gas in superbubbles following the explosion of supernovae. 


Our results show roughly $10-20\%$ of the total {\HI} mass to be extra-planar. A similar amount of extra-planar gas ($\sim 10\%$) was found for NGC 2403 by \citet{fraternalietal02} using a similar Gaussian profile fitting and extraction to what is used in this study. Interestingly, both NGC 2403 and NGC 4559 also have similar star formation rates of $1.0$ M$_{\odot}$ yr$^{-1}$ and 
$1.1$ M$_{\odot}$ yr$^{-1}$, respectively \citep{sandersetal03}. Additionally, NGC 3198 was found to house $\sim15\%$ of its total {\HI} mass as extra-planar gas, with a SFR of $0.61$ M$_{\odot}$ yr$^{-1}$ \citep{gentileetal13}. The amount of extra-planar gas in NGC 3198 comes from a tilted ring model analysis, in which $15\%$ of the total {\HI} mass is used in a thick disk component. Since some of that thick disk is superimposed on the thin disk, the amount of extra-planar gas in NGC 3198 is likely closer to $\sim10\%$, as we find in NGC 4559. Indeed, even when star formation rate density (SFR/D$_{25}^2$) is compared, both NGC 2403 and NGC 4559 are similar, with star formation rate densities of $0.045$ and $0.042$ M$_{\odot}$ yr$^{-1}$ kpc$^{-2}$ \citep{sandersetal03}. NGC 3198 has a lower star formation rate density of $0.016$ M$_{\odot}$ yr$^{-1}$ kpc$^{-2}$ \citep{sandersetal03}, yet still has a substantial amount of extra-planar gas \citep{gentileetal13}.

\subsection{Forbidden Gas and {\HI} Holes}

A striking amount of anomalous gas located in the
'forbidden' region of the position-velocity diagram along the
major-axis is present in the data cube of NGC 4559.  This feature was
also mentioned in B05, but the deeper HALOGAS data
makes it possible to study this region in more detail. The existence of other {\HI} holes is noted in this galaxy. However, we focus on this particular one due to its proximity to the forbidden gas feature. 

\begin{figure}
\centering
\includegraphics[scale=0.48]{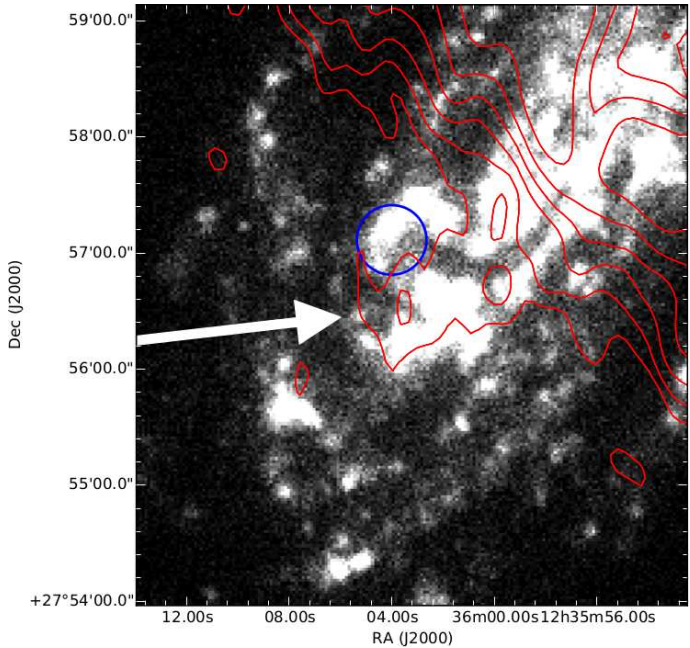}
\caption{Background image: GALEX FUV image; red contours: integrated $9$ channels in the full resolution cube containing forbidden gas emission. The blue circle outlines the position of the large {\HI} hole. The arrow points to the forbidden gas filament. The contour levels begin at $3$ times the rms noise in a single channel and increase in multiples of $3$.  }
\label{hole_chanplot}
\end{figure}

A study by \citet{boomsmaetal08} found that most of the {\HI} holes in NGC 6946 are related to extra-planar material expelled by star formation.
The previous study by B05 noted the potential
relationship between the forbidden velocity gas feature in NGC 4559 and a
nearby {\HI} hole. The same {\HI} hole is visible in the
full-resolution HALOGAS cube, as well. In Figure \ref{hole_chanplot}, we show
the sum of $9$ channels in the full resolution HALOGAS cube which showed the strongest forbidden emission, overlaid on the GALEX FUV image, and mark the locations of the {\HI} hole and forbidden velocity filament.
We estimate the location of the
center of this hole to be $\alpha=12^h36^m3.3^s$
$\delta=27^\circ57^m9.6^s$, which is in good agreement with B05. We
estimate the center of the forbidden gas feature to be located at
$\alpha=12^h36^m0.6^s$ $\delta=27^\circ56^m52.0^s$ -- this is
$\sim40\arcsec$ ($1.5$ kpc) away from the center of the hole on the sky.
Assuming the hole and the forbidden gas feature lie in the same plane on the sky, the vertical distance from the forbidden gas to the point on the plane directly below it would be $h=1.5 $ kpc $tan(67^\circ)=3.5$ kpc.

In order to explore the potential origins of this peculiar feature, we
extracted it from each channel in which it is present in the
$30\arcsec$ smoothed cube. Emission was found in 13 channels between
$817.78$ km s$^{-1}$ and $867.20$ km s$^{-1}$. The total mass of the
feature is $\sim 1.4\times 10^6$ M$_{\odot}$.

The proximity and
orientation of the forbidden gas filament suggests it could contain gas that may
have once filled the hole. Assuming the bulk motion of the gas in the forbidden gas filament can
be characterized by the velocity of the central channel of its
emission, we estimate the relative velocity of the filament to the gas
in its spatial surroundings. The central channel of the filament's
emission is at $842.5$ km s$^{-1}$, and the channel containing most
of the gas spatially coincident with the feature is at $719$ km
s$^{-1}$, ie. the bulk velocity of the gas around the filament has a velocity of $\sim 123$ km s$^{-1}$ relative to the hole. 
We estimate the amount of kinetic energy the forbidden feature would have moving directly vertically from the plane of the galaxy via $K=1/2 mv^2$. Using the relative velocity of the feature corrected for inclination, this energy would be $\sim1.8\times10^{53}$ erg. 

We estimate the potential energy required to move a parcel of gas to the height of the feature assuming it is located directly above the plane of the disk ($3.5$ kpc). We use Equation 4 of \citet{howkandsavage97} for the potential energy of extra-planar clouds above a galaxy disk. We use the above tabulated cloud mass and height above the plane, assume the stellar disk scale height is the model thin disk scale height of $\sim 200$ pc, and assume the mid-plane mass density is that of the Solar Neighborhood: $\rho_0 =0.185$ M$_{\odot}$ pc$^{-3}$ \citep{bahcall84}. We find the corresponding potential energy of the forbidden gas feature is $2 \times 10^{53}$ erg s$^{-1}$. The total energy, kinetic plus potential, required to move this gas is $\sim4.0\times10^{53}$ erg, or the energy of $\sim2000$ supernovae of energy $10^{51}$ erg, assuming $20\%$ efficiency. This efficiency and energy requirement is reasonable for a superbubble that arose from many supernovae. \citet{fraternalietal15} relate the Galactic HVC Complex C to ejection via star formation processes within the Milky Way. This complex has a hydrogen mass and relative velocity both roughly twice as large as seen in the forbidden gas feature of NGC 4559. \citet{fraternalietal15} estimate the number of supernovae required to eject the complex to be $\sim 1 \times 10^4$ supernovae, corresponding to a typical star formation rate density of a star forming region in the Milky Way disk. Thus, it is plausible that this feature in NGC 4559 was once part of the disk, but was ejected through star formation processes. 

If our assumptions about the vertical separation of the forbidden gas from the disk were incorrect, the energy calculation would change somewhat. The kinetic energy would be unchanged. However, the potential energy would decrease if the gas were actually closer to the disk, further bolstering our above conclusion. If the gas were actually located further from the disk, it would require more supernovae to potentially eject the forbidden gas feature. If the gas were located $7$ kpc above the disk, this would increase our potential energy calculation by a factor of $5$, which would not be enough to change the above conclusion. Of course, beyond some height, the potential energy required would begin to become unreasonably high. But, the greater the height of the gas, the less likely it is to show the smooth kinematic connection to permitted velocity gas in the major axis p-v diagram.

We can explore whether or not the forbidden gas is an outflow feature or infalling from purely geometric arguments. Both the hole and the forbidden gas feature are located near the major axis in the approaching side of the galaxy and we assume that the spiral arms seen in the GALEX image are trailing arms (see Figure \ref{hole_chanplot}), indicative of counter-clockwise rotation. If the arms are indeed trailing arms, then the SW side is the near side. 
If we assume that the spatial separation between the hole and the forbidden gas feature is mostly along the $z$-axis, then the feature must be located \textit{on the far side of} the disk, and is therefore outflowing due to its positive heliocentric velocity. If it is inflowing and on the near side of the disk, then the feature cannot be located above the hole, but would instead be ``ahead" of the hole in azimuth, which cannot be due to a lag. Therefore, if the feature is inflowing, it is either not related to the hole, or was launched from the hole at a large angle, which is not likely. Since we see this forbidden gas feature as a smooth connection to the extra-planar gas signatures at permitted velocities, which we attribute to a galactic fountain, we believe the forbidden gas feature to most likely be an outflow on the far side of the disk. 

In an attempt to explain the presence of the forbidden gas feature,
numerous tilted ring models containing both radial and vertical
inflows and outflows were created with TiRiFiC. Models were created
containing varying strengths of these flows in the inner regions of
the thick and thin disks of the best-fitting thick disk
model. However, no combination of these effects made models containing
any distinct feature akin to the forbidden gas feature. Simple
kinematic changes to the tilted ring model are insufficient to match
the observed phenomenon. This may be due to the nature of tilted ring
fitting -- we attempted to model an isolated, non-axisymmetric structure, within rings that
extend through half of the angular extent of the entire galaxy. We note that \citet{fraternalibinney06} show that a random fountain can produce similar, centrally-located, forbidden gas features that also exist in NGC 2403.

\section{{\HI} Dwarf Galaxy}

\begin{figure}
\centering
\includegraphics[scale=0.4]{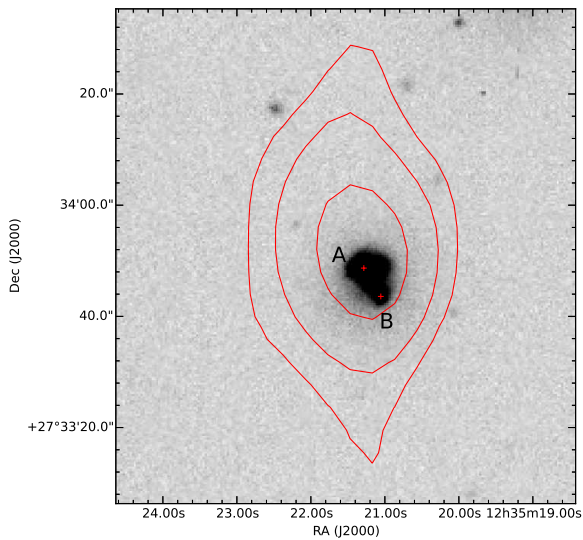}
\caption{SDSS g band overlaid with red {\HI} contours. The contour levels begin at $2.4\times10^{19}$ cm$^{-2}$ and increase in multiples of 2. The red crosses mark the positions of the two SDSS sources: point A is the north-eastern source, and point B is the south-western source. The object is a point-source in {\HI}, so the object's shape traces the beam.}
\label{possoverlay}
\end{figure}

A previously {\HI} undetected dwarf galaxy was found in the widest field HALOGAS data cube of NGC 4559. The center of this dwarf is located at $\alpha=12^h35^m21.335^s$, $\delta=27^\circ33^m46.68^s$, which puts it
at least $0.418^\circ$ ($\sim58$ kpc) away from the center of NGC 4559, if the objects are spatially aligned in the same plane. 
The heliocentric velocity of the dwarf is $\sim 1200$ km s$^{-1}$ and emission from the dwarf spans 9 channels from $1187-1224$ km s$^{-1}$, in total.
This also makes the dwarf well outside the field of view of the H$\alpha$ image. 
Velocities are computed in the optical definition.
The total {\HI} flux of this feature was found to be $27.1$ mJy, corresponding to a total
{\HI} mass of $\sim4\times10^{5}$ M$_{\odot}$, assuming a distance to the dwarf of $7.9$ Mpc. We show the {\HI} dwarf in Figure \ref{possoverlay}, overlaid on an SDSS g band image. 

We attempted to determine whether the dwarf is, in fact, a bound companion to NGC 4559. 
The circular velocity of NGC 4559 is $\sim 130$ km s$^{-1}$, as seen in the {\HI} rotation curve modelled in this study. We assume a typical halo mass for a galaxy at the rotation velocity of NGC 4559 by inverting Equation 8 in \citet{klypinetal11}. That study found an empirical relation between circular velocity and halo mass, assuming NFW dark matter density profiles using the Bolshoi simulation. The virial mass we find is $\sim 2.8 \times 10^{11}$ M$_{\odot}$, assuming $h=0.70$ also from the Bolshoi simulation \citep{klypinetal11}. We find the virial mass of a Milky Way-sized halo to be $\sim1.5 \times 10^{12}$ M$_{\odot}$ using that same relation. Since R$_{vir} \propto$ M$_{vir}^{1/3}$, we estimate the virial radius of NGC 4559 to be $\sim 170$ kpc. Even if the dwarf is not in the same plane as NGC 4559, it is still likely well within the virial radius. 

We estimate the escape velocity of the dwarf using Equation 2-192 in \citet{binneytremaine}. We estimate $r_\star \sim 25$ kpc, the maximum radius where the circular velocity is constant. Since the halo extends further than the {\HI} emission, this value is a lower limit, and thus makes the escape velocity estimate also a lower limit. In this calculation, we also use the circular velocity estimate from the tilted ring modelling ($v_{c}\sim 130$ km s$^{-1}$), and the assumed distance to the dwarf of $58$ kpc. This yields an escape velocity of $v_{\mathrm{esc}}\sim 121$ km s$^{-1}$. This value is much lower than the $390$ km s$^{-1}$ difference in systemic velocity of NGC 4559 and the central velocity of the dwarf. Though we cannot measure the exact distance where the rotation curve ceases to be flat, this would have to occur at a radial distance of $\sim 261$ kpc for the escape velocity to match the velocity of the dwarf. Since it is unlikely that the rotation curve is flat to such an extreme distance, we conclude that the dwarf is unbound. 

\begin{figure}
\includegraphics[scale=0.25]{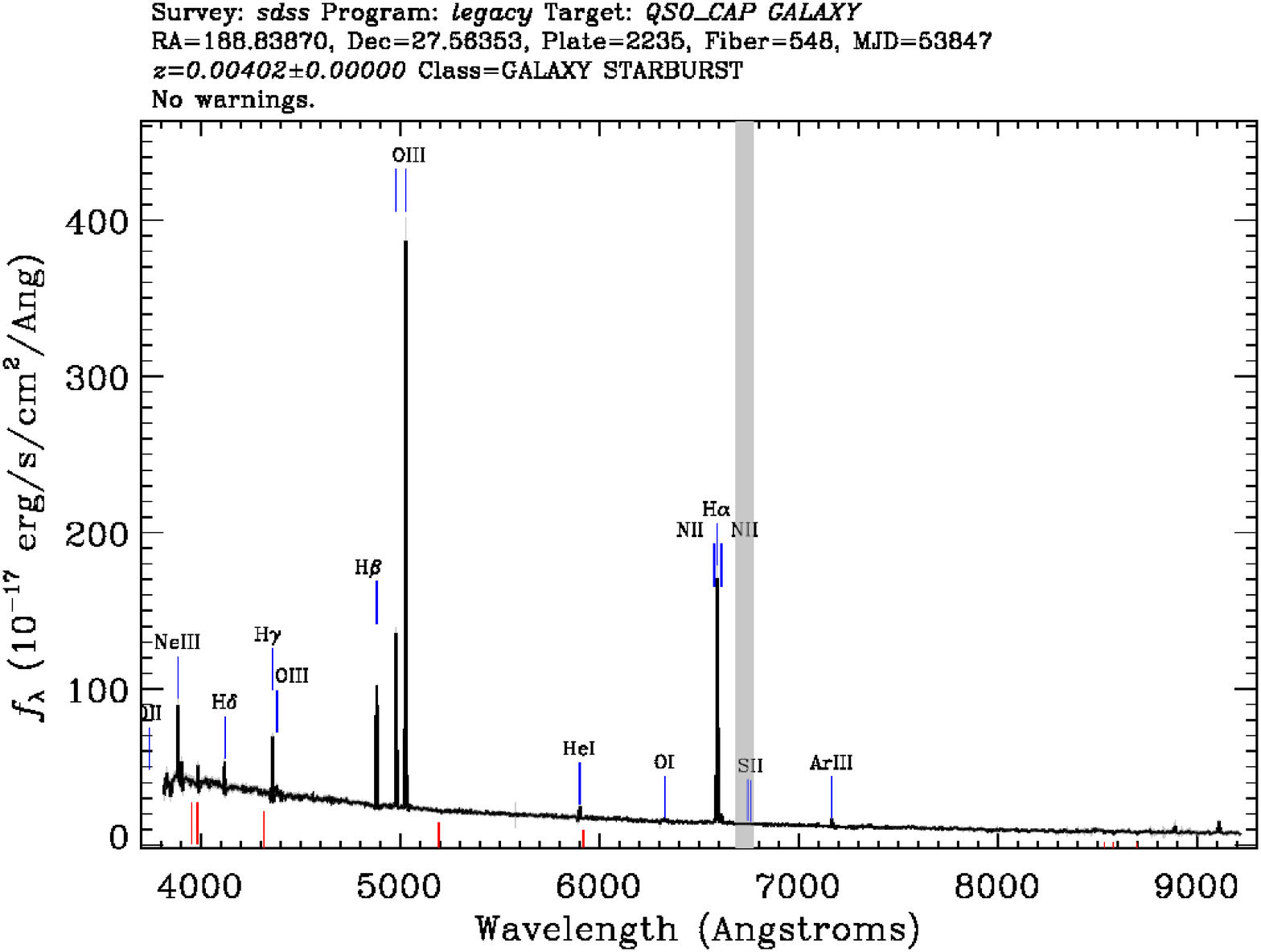} 
\includegraphics[scale=0.25]{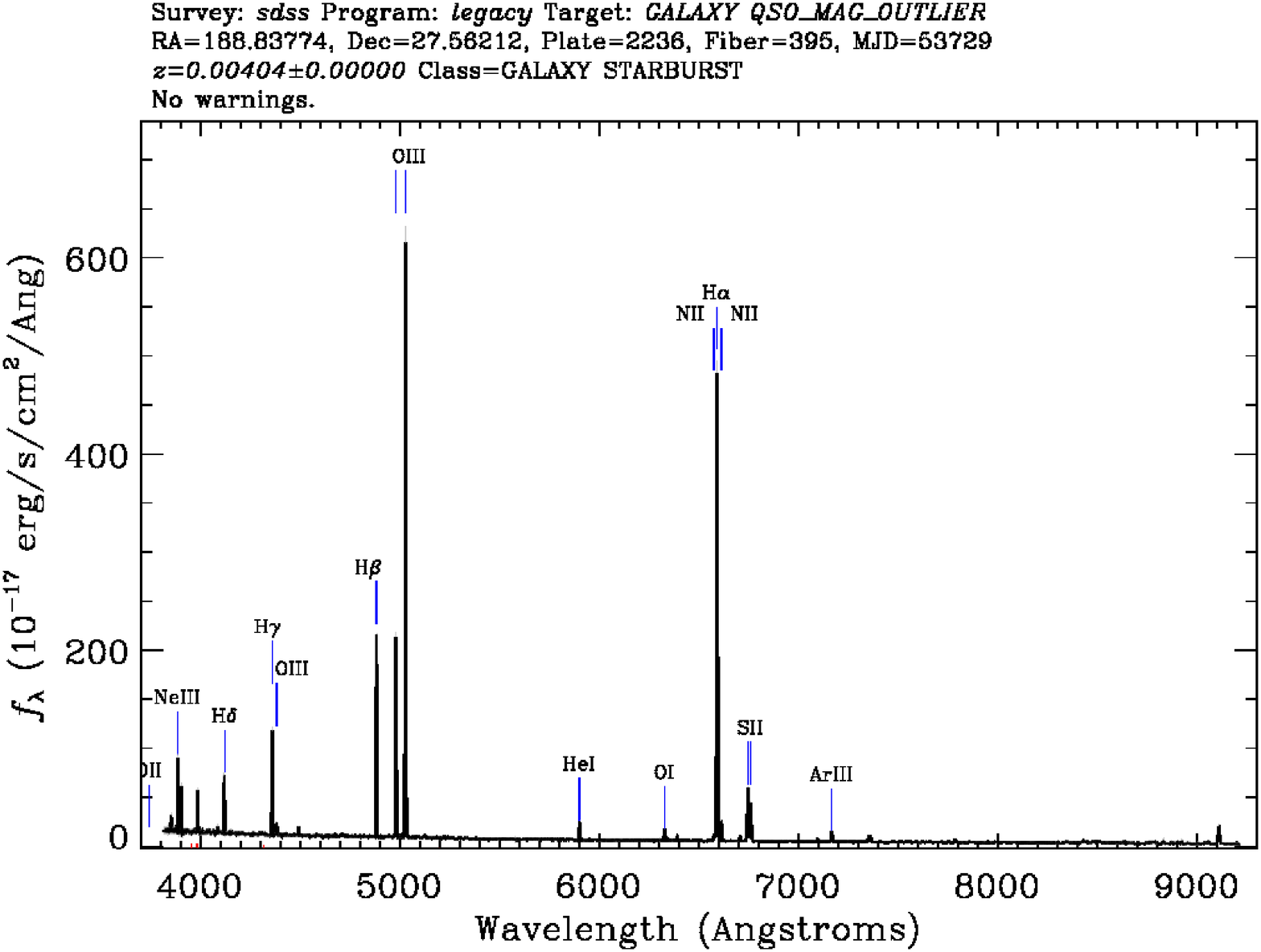}
\caption{SDSS sectra of the two regions associated with the BCD galaxy. The top panel is of the north-eastern region (A), and the bottom panel is of the south-western region (B). The vertical gray band encompasses masked pixels in the spectrograph. The blue and red lines point out the locations of possible emission and stellar absorption lines, respectively.  }
\label{companionspectra}
\end{figure}

Two optical counterparts for this object were found in the Sloan
Digital Sky Survey (SDSS, \citealt{eisensteinetal11}) data release 12
database. In Figure \ref{possoverlay} we show a SDSS g Band image with the HALOGAS {\HI} contours overlaid in red, showing the locations of the SDSS counterparts and the source in {\HI}.
These two objects are separated by $\sim5\arcsec$ and appear to be two merging objects. The SDSS spectra for these two regions show
the objects to both reside at $z=0.004$, which corresponds to a
heliocentric velocity of $1198$ km s$^{-1}$, which is effectively identical to
 the velocity of the dwarf in {\HI}. The spectra of these objects
can be seen in Figure \ref{companionspectra}. They
both show large OIII/H$\beta$ ratios, implying the existence of high
excitation HII regions, and the existence of young stars. We conclude that the dwarf is actually two merging Blue Compact Dwarf (BCD) galaxies.

The SDSS quoted u and g band magnitudes for the north-eastern object are $17.12$ and $16.67$, and those of the south-western object are $17.22$ and $16.79$. Using the assumed distance of $7.9$ Mpc, the corresponding absolute magnitudes are M$_{\mathrm{u}}=-12.4$ and M$_{\mathrm{g}}=-12.8$ for the north-eastern object, and M$_{\mathrm{u}}=-12.3$ and M$_{\mathrm{g}}=-12.7$ for the south-western object. We converted these magnitudes to an equivalent B band magnitude, then calculated the luminosity of the objects assuming they are at the same distance as NGC 4559. The dwarves have an {\HI} mass to blue light ratio of $0.18$. The optical size of the companion is $\sim5\arcsec$ or $191.5$ pc, assuming the object is at the distance of the galaxy. This {\HI} mass to blue light ratio is low, but within the values obtained by \citet{huchtmeieretal07} -- an Effelsberg {\HI} study of $69$ BCD galaxies. The combined {\HI} mass of these dwarves is also of the same order of magnitude to that of $12$ dwarf galaxies in the local group as described in \citet{mcconnachieetal12}. Of these 12, only 4 are within $200$ kpc of their parent galaxy. Also of these 12, 5 have M$_{\mathrm{V}}$ between $-14$ and $-11$. 

\section{Discussion and Conclusions}

Our analysis of the extra-planar {\HI} in NGC 4559 can
be compared to other moderately inclined spiral galaxies in the
HALOGAS sample. In a similar study by \citet{gentileetal13} of NGC
3198, a lagging extra-planar {\HI} component was discovered,
containing $\sim15\%$ of the total {\HI} mass of that galaxy. The
extra-planar {\HI} in NGC 3198 was found to be characterized by a
thick disk scale-height of $\sim3$ kpc with a vertical lag of $7-15$
km s$^{-1}$ kpc$^{-1}$. These values are very similar to what was
found in NGC 4559.

NGC 2403 is a very similar galaxy to NGC 4559 in morphology and star
formation characteristics, and was studied in {\HI} with the Very
Large Array by \citet{fraternalietal02}. NGC 2403 has a star formation
rate of $1$ M$_{\odot}$ yr$^{-1}$ \citep{sandersetal03} and a rotation
velocity of $122$ km s$^{-1}$. \citet{fraternalietal02} found NGC
2403 to contain an extra-planar {\HI} component containing $\sim10\%$
of the total {\HI} mass of that galaxy. That study found the
extra-planar {\HI} to be lagging the disk's rotation velocity by
$25-50$ km s$^{-1}$.

A recent study by \citet{debloketal14} of NGC 4414 found that only
about $\sim4\%$ of the total {\HI} mass is in extra-planar
gas. However, due to the disturbed nature of that galaxy's halo, that
number is difficult to constrain. Analysis of the inner disk of NGC
4414 show that galaxy has likely experienced a recent interaction with
a dwarf galaxy, which may account for its large star formation rate of
$4.2$ M$_{\odot}/yr$.

Although the characteristics of the extra-planar {\HI} in NGC 4559
seem to fit into the overall nearby moderately-inclined spiral galaxy
picture, it is difficult to say whether or not lagging extra-planar
{\HI} is ubiquitous or extraordinary throughout the nearby
universe. Furthermore, is extra-planar lagging {\HI} always seen to be
likely caused by the Galactic Fountain mechanism? Greater
understanding of this problem can be obtained through further study of
the entire HALOGAS sample, both edge-on and moderately inclined
galaxies. The next generation of radio telescopes, including the
Square Kilometre Array, will answer these questions in the future.

We used the deep 21 cm HALOGAS observations of NGC 4559 to expand upon
the work done by \citet{barbierietal05} in characterizing diffuse
extra-planar and anomalous characteristics in the {\HI} distribution
of that galaxy. We created detailed three dimensional, tilted ring
models of that galaxy's {\HI}. We confirm B05, in that a model containing only a
$\sim200$ pc thin disk cannot produce a fit to the faint extra-planar lagging gas signatures seen in this galaxy. We create an expanded model
containing a thick disk comprised of 20\% of the total {\HI} of the
thin disk model, whose thick disk component extends vertically to a
scale height of $2$ kpc. We constrain the magnitude of the gradient in rotation velocity 
with height in a simple thick disk model to be $\sim 13$ km s$^{-1}$ kpc$^{-1}$ in the approaching side and $\sim 6.5$ km s$^{-1}$ kpc$^{-1}$ in the receding side. In the fine-tuned model, we find that a lag of $\sim 13$ km s$^{-1}$ kpc$^{-1}$ in both halves, but with a cutoff in the receding half near $R_{25}$ is an improved match to the data. This measurement of the lag magnitude was not previously done in B05, where a separate rotation curve was used for the thick disk. 

We use a Gaussian line profile fitting technique to extract the
anomalously rotating extra-planar gas from the normally rotating
disk. In this technique we find that $\sim 10\%$ of the total {\HI}
mass is extra-planar. Also, the extra-planar gas is localized to the
inner star-forming regions of the galaxy, again suggesting a bulk of
this gas is of galactic fountain origin.

We analyze the spatial locations of total and extra-planar {\HI} in relation to H$\alpha$, emission seen from young stars as a tracer for active star formation. We find that extra-planar {\HI} traces regions of star formation, leading us to conclude that most of the extra-planar {\HI} seen is from in-situ star formation, ie. a galaxy-wide galactic fountain. 

To further build on the work of B05, we extracted the emission from a filament of {\HI}
located in the kinematically forbidden region of the
position-velocity diagram along the major axis. We find that the
feature contains $1.4$ $\times 10^6$ M$_\odot$ of {\HI}. Energy
estimates of the feature require $\sim2000$ supernova to move
the gas, which is consistent with a superbubble or other in-situ processes due to star formation. The remarkable proximity of this feature to a large {\HI} hole is difficult to ignore, but no irrefutable evidence tying the two together was found. Furthermore, the feature extends into the extra-planar gas signatures quite smoothly in the position-velocity diagram along the major axis, further pointing to the filament originating inside of the normally rotating disk, and was also expelled through star formation.

We analyze a merger of two BCD galaxies, previously unobserved in {\HI}, located $\sim0.4^{\circ}$ from the center of NGC 4559. The BCD galaxies contain $\sim4\times10^{5}$ M$_{\odot}$ of {\HI} and contain two spatially tight counterpart sources in SDSS. We conclude the objects are merging BCD galaxies due to a low {\HI} mass to blue light ratio of $0.18$ and spectra largely indicative of {\HII} regions. 

\acknowledgements
This material is based upon work supported by the National Science Foundation Graduate Research Fellowship under Grant No. 127229 to CJV. This material is also based on work partially supported by the National Science Foundation under Grant No. AST-1616513 to RJR and Grant Nos. AST-0908126 and AST-1615594 to RAMW. This project has received funding from the European Research Council (ERC) under the European Union's Horizon 2020 research and innovation programme (grant agreement No 679627). We would like to thank the anonymous referee for insightful and helpful comments that lead to the overall improvement of this work.

\clearpage

\newpage
\bibliography{refs_NGC4559}
\bibliographystyle{aasjournal}
\end{document}